\documentclass[a4paper,11pt]{article}                                     %
\usepackage{graphicx}                                                     %
\usepackage{amssymb}                         
\def\Vec#1{\mbox{\boldmath $#1$}}                                         %
\begin{document}                                                          %
%
%
%
%
\begin{titlepage}                                                         %
%
%
%
%
\begin{flushright}                                                        %
KEK-TH-784 \\                                                             %
Oct.\,5, 2001 \\                                                          %
\medskip                                                                  %
hep-th/0110106                                                            %
\end{flushright}                                                          %
%
%
%
%
\begin{center}                                                            %
\vspace{35mm} 
{\Large                                                                   %
\textbf{$\Vec{E_6}$\,\ Matrix Model}                                      %
}\\                                                                       %
\vspace{25mm} 
{\large                                                                   %
Yuhi Ohwashi                                                              %
\footnote{E-mail: yuhi@post.kek.jp}                                       %
}\\                                                                       %
\vspace{10mm}           %
{\itshape                                                                 %
Department of Particle and Nuclear Physics,\\                             %
The Graduate University for Advanced Studies (GUAS),\\                    %
1-1 Oho, Tsukuba, Ibaraki 305-0801, Japan\\                               %
\medskip                                                                  %
Theoretical Physics Group, KEK                                            %
}\\                                                                       %
\vspace{10mm}           %
\end{center}                                                              %
%
%
%
%
%
%
%
%
\begin{abstract}                                           %
We consider a new matrix model based on the simply connected {\itshape compact} exceptional Lie group $E_6$. A matrix Chern-Simons theory is directly derived from the invariant on $E_6$. It is stated that the similar argument as Smolin \cite{Smolin:2001wc} which derives an effective action of the matrix string type can also be held in our model. An important difference is that our model has twice as many degrees of freedom as Smolin's model has. One way to introduce the cosmological term is the compactification on directions. It is of great interest that the properties of the product space $\Vec{\mathfrak{J}^c} \times \Vec{\mathcal{G}}$, in which the degrees of freedom of our model live, are very similar to those of the physical Hilbert space.
%
\end{abstract}                                                            %
%
%
%
%
\end{titlepage}                                                           %
%
%
%
%
%
%
%
%
%
%
%
%
\section{Introduction}

In this paper, we construct a new matrix model which has a {\itshape compact} $E_6$ symmetry and a Chern-Simons like structure. The definition of $E_6$ itself (i.e. the {\itshape cubic form}) is adopted for the construction of the action. The resulting theory has a Chern-Simons term in the action as in the case of type $F_4$ (i.e. the {\itshape trilinear form}) \cite{Smolin:2001wc}. The {\itshape compactness} of $E_6$ derives {\itshape the postulate of positive definite metric} of our model.

Today particle physics except for gravity is well described by the standard model. However, gravity cannot be quantized in the same method because we cannot renormalize it. Therefore the main problem of current particle physics is to establish a consistent quantum theory which contains both the standard model and gravity. Under these circumstances, the most hopeful and popular candidate is the string theory.

The reason to favor the string theory is its wonderful nature. We can give as concrete examples that the theory has no ultraviolet divergence and includes gravitational field as well as matter and gauge fields automatically. However, due to the infinite ground states, this theory has no capability to predict; therefore we cannot answer why the standard model emerges. On the other hand it is possible to consider that this problem is the problem in the framework of perturbative formulation of the theory, because the completed region of the string theory is only the perturbative region. So if the non-perturbative formulation of the theory is accomplished, it is quite likely that this problem is resolved. Of course, it is pure speculation, but it seems quite probable that the non-perturbative effects turn infinite ground states into single one. In recent years, some kinds of non-perturbative effects of the string theory were investigated using the concepts like D-brane, duality, and M-theory. However these are not constructive definitions of the string theory as yet, but attempts to understand the non-perturbative effects along the line of the perturbation theory.

What must not be forgotten is that one theory never finish before the non-perturbative formulation is completed. One of candidates for the non-perturbative formulation of the string theory at present is the string field theory. Although a considerable number of studies have been conducted on these theories, the only successful string field theories so far are the ones formulated in the light-cone gauge. So it is not clear whether we can extract some essential information of the non-perturbative effects. Another candidate is what is called the matrix model. With the advent of the BFSS model \cite{Banks:1997vh}\footnote{The action itself has been introduced before by B.de Wit, J.Hoppe and H.Nicolai \cite{deWit:1988ig}.} as a starter, many proposals (e.g.\cite{Banks:1997vh,Ishibashi:1997xs,Dijkgraaf:1997vv,Smolin:2000kc,Smolin:2000fr,Azuma:2001re,Smolin:2001wc}) have been being made since. The common idea of these models is that they reproduce sting or membrane theory in the large-$N$ limit. In a sense the matrix model is similar to the lattice gauge theory, which is the non-perturbative formulation of the field theory, in that they can be analyzed using numerical simulation. Therefore it is reasonable to suppose that we will develop current matrix models a little further and find the true model.

A virtue of the matrix model is that it has a possibility of putting an interpretation on the space-time itself. However, some important questions such as ``what would be the real mechanism to realize the 4-dimensional world from the 10(or 11)-dimensional universe'' and ``how is the diffeomorphism introduced into the theory'' remain unsettled. One of them is the question of background independence. Consider the IKKT model \cite{Ishibashi:1997xs} for example. This model has an $SO(10) \times SU(N)$ symmetry, and this is just a symmetry like {\itshape some theory} was expanded around the flat background. Therefore we cannot deny the existence of different matrix model whose expansion around a special background gets the IKKT model. On this point Smolin proposed a new type of matrix model \cite{Smolin:2000kc} in which the action is cubic in matrices. Matrices are built from the super Lie algebra $osp(1|32;\mathbf{R})$, and one multiplet is pushed into a single supermatrix. Smolin's conjecture is that the expansions around different backgrounds of the $osp(1|32;\mathbf{R})$ matrix model will reduce to the BFSS or IKKT model. However, as far as the IKKT model is concerned, the theory made from Smolin's way dose not reproduce the supersymmetry of the IKKT model. That is, indeed the 10-dimensionality is realized, but the {\itshape half} of supersymmetry required by the IKKT model cannot be held. Anyway, the model described by a single matrix alone is very attractive, and Smolin's courageous attempt demonstrated one concrete possibility. Afterwards, ``how 10-dimensional IKKT model can be embedded in 11-dimensional Smolin's models'' were reconsidered in Ref.\cite{Azuma:2001re}. Those arguments were proceeded using group theoretical study on the supersymmetry of the IKKT model. In addition to those, one view on the diffeomorphism in the matrix model, a different $u(1|16,16)$ model from Smolin's, and some kind of {\itshape topological} effective action derived using Wigner-In\"on\"u contraction were also discussed.

Moreover, as Smolin's $u(1|16,16)$ model \cite{Smolin:2000fr} has demonstrated, the matrix models are not irrelevant to the {\itshape loop quantum gravity} which is another approach to the Theory of Everything. Furthermore, it was pointed out in Ref.\cite{Smolin:2001wc} that the matrix string theory \cite{Dijkgraaf:1997vv} has a connection with the matrix model based on the exceptional Jordan algebra $\Vec{\mathfrak{J}}$, while B.Kim and A.Schwarz have discussed in Ref.\cite{Kim:2001up}\footnote{The author learned of the existence of this study from Smolin's paper Ref.\cite{Smolin:2001wc}.} a tie-in between the IKKT model and the Jordan algebra $\Vec{\mathfrak{j}}$ with its spinor representation. For these reasons, doing research on extended matrix model is very interesting and important. Over and above, we should not overlook the fact that several approaches which are very similar to the matrix model have been pursued by other fields. We can take Ref.\cite{Chamseddine:1997zu} from the non-commutative geometry, Ref.\cite{Klimcik:1998mg} from the fuzzy sphere, and Ref.\cite{Kawamoto:1999gi} from the simplicial lattice for example. It might be inferred from these circumstantial evidence that the attempt to renounce the space-time as a {\itshape continuum} holds one important key to the future progress of physics. It seems at least that there is no need to relate the matrix model to the string theory alone.

For these purposes, we consider a new matrix model based on the simply connected compact exceptional Lie group $E_6$ in this short article. This paper is organized as follows. In the next section we briefly review Smolin's matrix model based on the groups of type $F_4$. After that, in section 3, our model is presented. The action of the model is constructed from the {\itshape cubic form} which is the {\itshape invariant} on $E_6$ mapping. This action is an essentially complex action. Of course if one wants, one may take up only real part of the action; however, there are some circumstantial evidence where it is essential for the theory including gravity to employ complex variables. Ashtekar variables make the constraints of the canonical formalism of general relativity quite easy \cite{Ashtekar:1986yd,Ashtekar:1987gu}. Actually the chiral action is a complex action whose real part agrees with Palatini action. Besides, the $u(1|16,16)$ model have also complex action because of the $33th$ component which is pure imaginary. As in the loop quantum gravity, we may be able to impose the {\itshape reality condition} on our model. In addition, it is possible that the action expanded around the vacuum, which is a specific background of the model, gets on-shell real. In this paper, therefore, we do not restrict the action to real part only. We proceed with arguments using complex action. In section 4, we discuss symmetries which our model possesses. Supersymmetry has a deep connection with {\itshape cycle mapping} $\mathcal{P}$. In section 5, we investigate constraints which are imposed on our model. The reason why the model is a constrained system is that the group we consider here is the {\itshape compact} $E_6$ group of the groups of type $E_6$. The resulting conditions are quite similar to what we postulate as fundamental properties of inner products of the physical Hilbert space. In section 6, we follow Smolin. We investigate one of the classical solutions of our model, and compactify the theory on the three directions. Our model has twice as many degrees of freedom as Smolin's model has because we are considering $E_6$ instead of $F_4$. However, we can have the similar argument as Smolin \cite{Smolin:2001wc} which derives an effective action of the matrix string type. The paper closes with brief discussions in section 7. In appendix, the author put together an elementary knowledge about the complex Graves-Cayley algebra $\Vec{\mathfrak{C}^c}$ and the complex exceptional Jordan algebra $\Vec{\mathfrak{J}^c}$ which are needed for this paper. However, if the reader has a great interest in the exceptional linear Lie groups, see, in particular, a series of excellent reports presented by I.Yokota \cite{Yokota:1990e6,Yokota:1990e7,Yokota:1991e8} one time.
\subsubsection*{Why $\Vec{E_6}$ ?}
On the end of this section, let us examine the reason why $E_6$ is better than the groups of type $F_4$. Although, as early as 1983, T.Kugo and P.Townsend referred to the relevance between physics and division algebras \cite{Kugo:1983bn}, the reason why the concern with $F_4$ has been growing is that the critical dimension of the string theory is 10 dimensions. The exceptional Jordan algebra $\Vec{\mathfrak{J}}$ is a 27-dimensional $\mathbf{R}$-vector space. This space can be classified into three main parts. One is the Jordan algebra $\Vec{\mathfrak{j}}$ which is a 10-dimensional $\mathbf{R}$-vector space. The others are the part of 16 dimensions which is related to the spinors and the extra 1 dimension. In brief, what are expected as the degrees of freedom of the Theory of Everything are all involved in this $\Vec{\mathfrak{J}}$. The extra 1 dimension may account for the M-theory. Moreover, an important point to emphasize is the fact that the groups of type $F_4$ have some definite geometrical interpretations. For example, $F_4$ is a subgroup of projective transformation group of Graves-Cayley projective geometry $\Vec{\mathfrak{C}} P$ which corresponds to the {\itshape elliptic} non-Euclidean geometry; and $F_{4(-20)}$ is a subgroup of projective transformation group of Graves-Cayley projective geometry which corresponds to the {\itshape hyperbolic} non-Euclidean geometry. For these reasons, the groups of type $F_4$ are very attractive to us, but there is one flaw in these groups. That is the fact that elements of $\Vec{\mathfrak{J}}$ do not have an imaginary unit `$i$'. It is true the matrix model is one of candidates for the unified theory. Now may be the state of affairs in the model building, but we will have to account for the standard model someday. Fermions which appear in the standard model are Weyl spinors, namely {\itshape complex} spinors. If the standard model were described by using Majorana spinors only, $F_4$ might be the underlying symmetry of the universe. However, the actual world requires {\itshape complex} fermions without doubt. In the usual string theory, the real $\gamma$-matrices in 10 dimensions can be decomposed by direct products of Pauli matrices, so that the appropriate compactification can change 10-dimensional Majorana fermions into 4-dimensional Weyl fermions. In the approaches using Jordan type algebras, however, it needs care because the elements are non-associative which have no matrix representations. We cannot use above usual trick. Let us take one concrete example. In the case of $osp(1|32;\mathbf{R})$ supermatrix models the bosonic part can be expanded in terms of 11-dimensional $\gamma$-matrices with rank 1,2 and 5 \cite{Smolin:2000kc,Azuma:2001re}, and in the case of $u(1|16,16)$ models the bosonic part can be expanded in terms of 11-dimensional $\gamma$-matrices with rank 0,1,2,3,4 and 5 \cite{Smolin:2000fr,Azuma:2001re}. These super Lie algebras contain the $\gamma$-matrices clearly, but the Jordan type algebras do not have the usual (associative) Clifford algebra trivially. Therefore, the complex structure needs to be introduced into the theory from the beginning.

This is the reason why we consider $E_6$ is better than the simply connected compact exceptional Lie group $F_4$. Of course, the group $F_4{}\Vec{^c}$ is still available, but the model based on $F_4{}\Vec{^c}$ is merely the complexification of the Smolin's model. So we attempt to use the {\itshape compact} $E_6$ group. The complex exceptional Jordan algebra $\Vec{\mathfrak{J}^c}$ is the complexification of the exceptional Jordan algebra $\Vec{\mathfrak{J}}$. One must not confuse the complexification of the exceptional Lie group with that of the exceptional Jordan algebra. Writing down some definitions of the exceptional Lie groups is very informative for the comparison.
\begin{eqnarray}
F_4 &=& \{ \alpha \in Iso_R ( \Vec{\mathfrak{J}} , \Vec{\mathfrak{J}} ) \, | \, \  tr ( \alpha A , \alpha B , \alpha C ) = tr ( A , B , C ) \, , \, ( \alpha A , \alpha B ) = ( A , B ) \} \nonumber \\
F_4{}\Vec{^c} &=& \{ \alpha \in Iso_C ( \Vec{\mathfrak{J}^c} , \Vec{\mathfrak{J}^c} ) \, | \, \  tr ( \alpha X , \alpha Y , \alpha Z ) = tr ( X , Y , Z ) \, , \, ( \alpha X , \alpha Y ) = ( X , Y ) \} \nonumber
\end{eqnarray}
\begin{eqnarray}
E_{6(-26)} &=& \{ \alpha \in Iso_R ( \Vec{\mathfrak{J}} , \Vec{\mathfrak{J}} ) \, | \, \  ( \alpha A , \alpha B , \alpha C ) = ( A , B , C ) \} \nonumber \\
E_6 &=& \{ \alpha \in Iso_C ( \Vec{\mathfrak{J}^c} , \Vec{\mathfrak{J}^c} ) \, | \, \  ( \alpha X , \alpha Y , \alpha Z ) = ( X , Y , Z ) \, , \, \langle \alpha X , \alpha Y \rangle = \langle X , Y \rangle \} \nonumber \\
E_6{}\Vec{^c} &=& \{ \alpha \in Iso_C ( \Vec{\mathfrak{J}^c} , \Vec{\mathfrak{J}^c} ) \, | \, \  ( \alpha X , \alpha Y , \alpha Z ) = ( X , Y , Z ) \} \nonumber
\end{eqnarray}
Here, $tr ( \ast , \ast , \ast )$ is the {\itshape trilinear form}, and $( \ast , \ast , \ast )$ is the {\itshape cubic form}. The cubic form is very different from the trilinear form, and their concrete forms are given in appendix.

Another virtue is that $E_6$ contains $Spin(10)$. This is never achieved by type $F_4$, so this is a very good point of our model. It is quite likely that other matrix models are reproduced via expansions around specific backgrounds of our model. In addition, $E_6$ is also interesting from the viewpoint of phenomenology.

On the other hand, the problem now arises:\ \ Naturally if we deal with {\itshape compact} $E_6$, the degrees of freedom of the theory double because of $\Vec{\mathfrak{J}^c}$. Although this problem always follows us as long as we handle {\itshape compact} $E_6$, even so, it seems that the benefit of the fact that we can introduce complex structure into the theory exceeds this trouble. Therefore we have to prepare a mechanism separately, which reduces the number of degrees of freedom by half. This is a future problem which needs to be asked.
%
%
%
%
\section{Smolin's matrix model based on the groups of type $F_4$}

In this section, we briefly review Smolin's matrix model based on the groups of type $F_4$ \cite{Smolin:2001wc}.

The action of Smolin's model is given by 
\begin{eqnarray}
S &=& \frac{k}{4 \pi} \  \  tr \Bigl( \  J^A \  , \  \mathcal{P} ( J^B ) \  , \  \mathcal{P}^2 ( J^C ) \  \Bigr) \  \  f_{ABC} \  \  \  ,
\end{eqnarray}
where $J^A$ are elements of real exceptional Jordan algebra $\Vec{\mathfrak{J}}$, \,$\mathcal{P} ( J )$ is the cycle mapping of $J$, \  $tr ( A , B , C ) \  \  ( A,B,C \in \Vec{\mathfrak{J}} )$ is the {\itshape trilinear form}, \,and $f_{ABC}$ are the structure constants of $\Vec{\mathcal{G}}$ which is a Lie algebra. Therefore the degrees of freedom of this model live in $\Vec{\mathfrak{J}} \times \Vec{\mathcal{G}}$.

The specific components of $J^A \in \Vec{\mathfrak{J}}$ are written as 
\begin{eqnarray}
J^A
&=&
\left(
      \begin{array}{ccc}
      z_1{}^A & \mathcal{O}_0{}^A & \bar{\mathcal{O}}_2{}^A \\
      \bar{\mathcal{O}}_0{}^A & z_2{}^A & \mathcal{O}_1{}^A \\
      \mathcal{O}_2{}^A & \bar{\mathcal{O}}_1{}^A & z_0{}^A
      \end{array}
\right) \\
& &
\quad\quad\quad z_I{}^A \in \mathbf{R} \ \quad \mathcal{O}_I{}^A \in \Vec{\mathfrak{C}} \quad\quad (I=0,1,2) \  , \nonumber
\end{eqnarray}
where $z_I{}^A$ are real numbers, and $\mathcal{O}_I{}^A$ are elements of Graves-Cayley algebra. 

In this model, following change of variables are required in order to make the theory Chern-Simons type, 
\begin{eqnarray}
x_0 \equiv z_1 + z_2 \  , \quad x_1 \equiv z_2 + z_0 \  , \quad x_2 \equiv z_0 + z_1 \  ,
\end{eqnarray}
and the effective theory of $T^3$ compactified theory of this model is expected to reproduce, at the one loop level, a theory related to the matrix string theory.
%
\section{The model}

We adopt a following definition for the simply connected {\itshape compact} exceptional Lie group $E_6$.
\begin{eqnarray}
E_6 = \{ \alpha \in Iso_C ( \Vec{\mathfrak{J}^c} , \Vec{\mathfrak{J}^c} ) \  | \  \  ( \alpha X , \alpha Y , \alpha Z ) = ( X , Y , Z ) \  , \  \langle \alpha X , \alpha Y \rangle = \langle X , Y \rangle \} \nonumber \\
\end{eqnarray}
The second condition $\langle \alpha X , \alpha Y \rangle = \langle X , Y \rangle$ is added to the definition to make the group {\itshape compact}.

We then define our theory by the following action,
\begin{eqnarray}
S &=& \Bigl( \  \mathcal{P}^2 ( \mathcal{M}^{[A} ) \  , \  \mathcal{P} ( \mathcal{M}^B ) \  , \  \mathcal{M}^{C]} \  \Bigr) \  \  f_{ABC} \  \  \  , \label{eq:E_6}
\end{eqnarray}
where $\mathcal{M}^A$ are elements of complex exceptional Jordan algebra $\Vec{\mathfrak{J}^c}$, \,$\mathcal{P} ( \mathcal{M} )$ is the cycle mapping of $\mathcal{M}$, \  \,$( X , Y , Z ) \  \  ( X,Y,Z \in \Vec{\mathfrak{J}^c} )$ is the {\itshape cubic form}, \,and $f_{ABC}$ are the structure constants of $\Vec{\mathcal{G}}$ which is a Lie algebra whose Lie group is {\itshape compact}. \  $[ \cdots ]$ denotes the {\itshape weight}-1 anti-symmetrization on indices.

The cubic form cannot be coupled to the structure constants as it is, because it is symmetric with respect to the interchange of fields. The cycle mapping is introduced in order to combine the cubic form with $f_{ABC}$. The way of constructing this action is basically the same as Smolin's case, but the {\itshape invariant} is different because the group we adopt is different from Smolin's. The {\itshape cubic form} is employed instead of the {\itshape trilinear form}. Of course, the exceptional Jordan algebra is not real $( \Vec{\mathfrak{J}} )$, but complex $( \Vec{\mathfrak{J}^c} )$. Therefore the degrees of freedom of our model live in $\Vec{\mathfrak{J}^c} \times \Vec{\mathcal{G}}$. One of the methods to introduce the cosmological term or the coupling constant into the theory is the compactification on directions. The physical dimensions can also be introduced via this compactification. Up till then, only the units of angle are in existence.

A beauty of this model is that there is no need to change variables intentionally in order to get a Chern-Simons type action.

The specific components of $\mathcal{M}^A \in \Vec{\mathfrak{J}^c}$ are written as follows, 
\begin{eqnarray}
\mathcal{M}^A
&=&
\left(
      \begin{array}{ccc}
      Q_1{}^A & \phi_3{}^A & \bar{\phi}_2{}^A \\
      \bar{\phi}_3{}^A & Q_2{}^A & \phi_1{}^A \\
      \phi_2{}^A & \bar{\phi}_1{}^A & Q_3{}^A
      \end{array}
\right)
+
i
\left(
      \begin{array}{ccc}
      P_1{}^A & \pi_3{}^A & \bar{\pi}_2{}^A \\
      \bar{\pi}_3{}^A & P_2{}^A & \pi_1{}^A \\
      \pi_2{}^A & \bar{\pi}_1{}^A & P_3{}^A
      \end{array}
\right) \\
& &
\quad\quad\quad Q_I{}^A, P_I{}^A \in \mathbf{R} \ \quad \phi_I{}^A, \pi_I{}^A \in \Vec{\mathfrak{C}} \quad\quad (I=1,2,3) \nonumber \\
&=&
\left(
      \begin{array}{ccc}
      \mathcal{A}_1{}^A & \Phi_3{}^A & \bar{\Phi}_2{}^A \\
      \bar{\Phi}_3{}^A & \mathcal{A}_2{}^A & \Phi_1{}^A \\
      \Phi_2{}^A & \bar{\Phi}_1{}^A & \mathcal{A}_3{}^A
      \end{array}
\right) \label{eq:components} \\
& &
\quad\quad\quad \mathcal{A}_I{}^A \in \mathbf{C} \ \quad \Phi_I{}^A \in \Vec{\mathfrak{C}^c} \quad\quad (I=1,2,3) \nonumber \\
&\equiv&
\mathcal{M}^A(\mathcal{A},\Phi) \  ,
\end{eqnarray}
where $\mathcal{A}_I{}^A$ are complex numbers, and $\Phi_I{}^A$ are elements of complex Graves-Cayley algebra.

In order to analyze the theory, it is necessary to decompose the action in terms of the variables defined in (\ref{eq:components}). This gives us 
\begin{eqnarray}
S &=& \frac{1}{4} \  f_{ABC} \  \epsilon^{IJK} \  \mathcal{A}_I{}^A \mathcal{A}_J{}^B \mathcal{A}_K{}^C + \frac{3}{2} \  f_{ABC} \  \epsilon^{IJK} \  \mathcal{A}_I{}^A ( \Phi_J{}^B , \Phi_K{}^C ) \nonumber \\
& & {} - 3 \  f_{ABC} \  \Vec{Re^c} ( \Phi_3{}^A \Phi_2{}^B \Phi_1{}^C ) + f_{ABC} \sum_{I=1}^{3} ( \  \Vec{Re^c} ( \Phi_I{}^A \Phi_I{}^B \Phi_I{}^C ) \  ) \\
&=& \frac{1}{4} \  f_{ABC} \  \epsilon^{IJK} \  \mathcal{A}_I{}^A \mathcal{A}_J{}^B \mathcal{A}_K{}^C + \frac{3}{4} \  f_{ABC} \  \epsilon^{IJK} \  \mathcal{A}_I{}^A ( \bar{\Phi}_J{}^B \Phi_K{}^C + \bar{\Phi}_K{}^C \Phi_J{}^B ) \nonumber \\
& & {} - \frac{3}{2} \  f_{ABC} \  ( \Phi_3{}^A (\Phi_2{}^B \Phi_1{}^C) + (\bar{\Phi}_1{}^C \bar{\Phi}_2{}^B) \bar{\Phi}_3{}^A ) \nonumber \\
& & {} + \frac{1}{2} \  f_{ABC} \sum_{I=1}^{3} ( \Phi_I{}^A (\Phi_I{}^B \Phi_I{}^C) + (\bar{\Phi}_I{}^C \bar{\Phi}_I{}^B) \bar{\Phi}_I{}^A ) \  ,
\end{eqnarray}
and depending on circumstances, it is helpful to rewrite these expressions using following relations.
\begin{eqnarray}
( \Phi_J{}^B , \Phi_K{}^C ) &=& \Vec{Re^c} ( \bar{\Phi}_J{}^B \Phi_K{}^C ) \\
f_{ABC} \sum_{I=1}^{3} ( \  \Vec{Re^c} ( \Phi_I{}^A \Phi_I{}^B \Phi_I{}^C ) \  ) &=& - f_{ABC} \  \sigma_{ijk} \sum_{I=1}^{3} ( \Phi_{iI}{}^A \Phi_{jI}{}^B \Phi_{kI}{}^C ) \\
f_{ABC} \  \Vec{Re^c} ( \Phi_3{}^A \Phi_2{}^B \Phi_1{}^C ) &=& - \frac{1}{6} \  f_{ABC} \  \epsilon^{IJK} \  \Vec{Re^c} ( \Phi_I{}^A \Phi_J{}^B \Phi_K{}^C ) \nonumber \\
& & {} \  - f_{ABC} \  \sigma_{ijk} \  \Phi_{i1}{}^A \Phi_{j2}{}^B \Phi_{k3}{}^C
\end{eqnarray}
Of course there is no kinetic term because we are considering the matrix model, but the first term of this action reproduces the Chern-Simons term. What has to be noticed is that this Chern-Simons type action is directly derived from the invariant on $E_6$. Therefore it is quite likely that one quality of the theory based on $E_6$ is {\itshape topological}.

Now, in order to get the equations of motion of this theory, we have to vary with respect to all fields. We first vary the action with respect to $\mathcal{A}_L{}^D$, and the resulting equations are the following.
\begin{eqnarray}
\frac{\delta S}{\delta \mathcal{A}_L{}^D} &=& \frac{3}{2} \  f_{ABD} \  \epsilon^{IJL} \  \Bigl( \  \frac{1}{2} \mathcal{A}_I{}^A \mathcal{A}_J{}^B + ( \Phi_I{}^A , \Phi_J{}^B ) \  \Bigr) \nonumber \\
&=& \frac{3}{2} \  f_{ABD} \  \epsilon^{IJL} \  \Bigl( \  \frac{1}{2} \mathcal{A}_I{}^A \mathcal{A}_J{}^B + \Phi_{0I}{}^A \Phi_{0J}{}^B + \Phi_{iI}{}^A \Phi_{iJ}{}^B \  \Bigr) \nonumber \\
&=& 0
\end{eqnarray}
Naturally, the equations of motion are obtained by differentiations with respect to the fields only, because there is no derivative term in the action. It is often convenient to rewrite these equations in terms of matrix form.
\begin{eqnarray}
\frac{\delta S}{\delta \mathcal{A}_L{}^D} \  \Vec{\tau}^L \mathbf{T}_D &=& - \frac{3}{2} \  \Bigl( \  \frac{1}{2} \mathcal{A}_I{}^A \mathcal{A}_J{}^B + \Phi_{0I}{}^A \Phi_{0J}{}^B + \Phi_{iI}{}^A \Phi_{iJ}{}^B \  \Bigr) \  [ \Vec{\tau}^I , \Vec{\tau}^J ] \  [ \mathbf{T}_A , \mathbf{T}_B ] \nonumber \\
&=& 0
\end{eqnarray}
The variations with respect to $\Phi_{0L}{}^D$ give us 
\begin{eqnarray}
\frac{\delta S}{\delta \Phi_{0L}{}^D} &=& \frac{3}{2} \  f_{ABD} \  \epsilon^{IJL} \  \Bigl( \   2 \mathcal{A}_I{}^A \Phi_{0J}{}^B + \Vec{Re^c} ( \Phi_I{}^A \Phi_J{}^B ) \  \Bigr) \nonumber \\
&=& \frac{3}{2} \  f_{ABD} \  \epsilon^{IJL} \  \Bigl( \  2 \mathcal{A}_I{}^A \Phi_{0J}{}^B + \Phi_{0I}{}^A \Phi_{0J}{}^B - \Phi_{iI}{}^A \Phi_{iJ}{}^B \  \Bigr) \nonumber \\
&=& 0 \  .
\end{eqnarray}
Similarly, we read off the following equations of motion from the variations with respect to $\Phi_{lL}{}^D \  (l=1,\cdots,7)$,
\begin{eqnarray}
\frac{\delta S}{\delta \Phi_{lL}{}^D} &=& 3 \  f_{ABD} \  \epsilon^{IJL} \  \Bigl( \  \mathcal{A}_I{}^A \Phi_{lJ}{}^B - \Phi_{0I}{}^A \Phi_{lJ}{}^B \  \Bigr) \nonumber \\
& & {} + 3 \  f_{ABD} \  \sigma_{ijl} \  \Bigl( \  \Phi_{i(L+1)}{}^A \Phi_{j(L+2)}{}^B - \Phi_{iL}{}^A \Phi_{jL}{}^B \  \Bigr) \nonumber \\
&=& 0 \  ,
\end{eqnarray}
where the index $L$ \,is $\Vec{mod \  3}$, \,and the summation convention is not used about this $L$.

We now introduce the following notations.
\begin{eqnarray}
\Vec{\mathcal{A}}_I = \mathcal{A}_I{}^A \  \mathbf{T}_A \  , \quad \Vec{\Phi}_{0I} = \Phi_{0I}{}^A \  \mathbf{T}_A \  , \quad \Vec{\Phi}_{iI} = \Phi_{iI}{}^A \  \mathbf{T}_A
\end{eqnarray}
These enable us to write preceding equations of motion as 
\begin{eqnarray}
\left\{
\begin{array}{r}
\epsilon^{IJL} \  \Bigl( \  \frac{1}{2} \  [ \Vec{\mathcal{A}}_I , \Vec{\mathcal{A}}_J ] + [ \Vec{\Phi}_{0I} , \Vec{\Phi}_{0J} ] + [ \Vec{\Phi}_{iI} , \Vec{\Phi}_{iJ} ] \  \Bigr) = 0 \\
\epsilon^{IJL} \  \Bigl( \  2 \  [ \Vec{\mathcal{A}}_I , \Vec{\Phi}_{0J} ] + [ \Vec{\Phi}_{0I} , \Vec{\Phi}_{0J} ] - [ \Vec{\Phi}_{iI} , \Vec{\Phi}_{iJ} ] \  \Bigr) = 0 \\
\epsilon^{IJL} \  \Bigl( \  [ \Vec{\mathcal{A}}_I , \Vec{\Phi}_{lJ} ] - [ \Vec{\Phi}_{0I} , \Vec{\Phi}_{lJ} ] \  \Bigr) + \sigma_{ijl} \  \Bigl( \  [ \Vec{\Phi}_{i(L+1)} , \Vec{\Phi}_{j(L+2)} ] - [ \Vec{\Phi}_{iL} , \Vec{\Phi}_{jL} ] \  \Bigr) = 0
\end{array}
\right. \  . \nonumber \\
\end{eqnarray}
Likewise, we obtain the following form for the action,
\begin{eqnarray}
S &=& - \frac{3i}{2} \  \epsilon^{IJK} \  tr \Bigl( \  \frac{1}{3} \Vec{\mathcal{A}}_I [ \Vec{\mathcal{A}}_J , \Vec{\mathcal{A}}_K ] + 2 \Vec{\Phi}_{\tilde{i}I} [ \Vec{\mathcal{A}}_J , \Vec{\Phi}_{\tilde{i}K} ] \  \Bigr) \nonumber \\
& & {} + 6i \  tr \Bigl( \  \Vec{\Phi}_{01} [ \Vec{\Phi}_{03} , \Vec{\Phi}_{02} ] - \Vec{\Phi}_{i1} [ \Vec{\Phi}_{03} , \Vec{\Phi}_{i2} ] - \Vec{\Phi}_{i1} [ \Vec{\Phi}_{i3} , \Vec{\Phi}_{02} ] - \Vec{\Phi}_{01} [ \Vec{\Phi}_{i3} , \Vec{\Phi}_{i2} ] \nonumber \\
& & {} \qquad\quad\quad - \sigma_{ijk} \  ( \Vec{\Phi}_{i1} [ \Vec{\Phi}_{j3} , \Vec{\Phi}_{k2} ] ) + \frac{1}{3} \sigma_{ijk} \sum_{I=1}^{3} ( \Vec{\Phi}_{iI} [ \Vec{\Phi}_{jI} , \Vec{\Phi}_{kI} ] ) \  \Bigr) \  , \label{eq:E_6decompose}
\end{eqnarray}
where $( \tilde{i} = 0,\cdots,7 )$ and $( i = 1,\cdots,7 )$.
%
\section{Symmetries of the model}

In this section, the symmetries of our model are described. First of all, the action is invariant under the following $E_6$ mapping as a result of our definition for the simply connected compact exceptional Lie group $E_6$.
\begin{eqnarray}
\lefteqn{\Bigl( \  \alpha \  \mathcal{P}^2 ( \mathcal{M}^{[A} ) \  , \  \alpha \  \mathcal{P} ( \mathcal{M}^B ) \  , \  \alpha \  \mathcal{M}^{C]} \  \Bigr) \  \  f_{ABC}}\hspace{1cm} \nonumber \\
&=& \Bigl( \  \mathcal{P}^2 ( \mathcal{M}^{[A} ) \  , \  \mathcal{P} ( \mathcal{M}^B ) \  , \  \mathcal{M}^{C]} \  \Bigr) \  \  f_{ABC} \\
&=& S \nonumber
\end{eqnarray}

Then, we have the gauge symmetry for the compact Lie group because the structure constants $f_{ABC}$ of $\Vec{\mathcal{G}}$ are coupled on to this action. We can take $\Vec{\mathcal{G}} = u(N)$ for example.

Moreover, as we can clearly see in (\ref{eq:E_6decompose}), there exist matrix translation symmetries with respect to the diagonal parts of the fields.

In addition to these, there is a particular symmetry which we call {\itshape cycle mapping} $\mathcal{P}$. This mapping is defined by the cyclic permutation with respect to the indices $I = 1, 2, 3$\,, and probably it belongs to the $F_4$. The action is invariant under the transformation $\mathcal{M}^A \mapsto \mathcal{P} ( \mathcal{M}^A )$.
\begin{eqnarray}
\lefteqn{\Bigl( \  \mathcal{P}^2 ( \mathcal{P} ( \mathcal{M}^{[A} ) ) \  , \  \mathcal{P} ( \mathcal{P} ( \mathcal{M}^B ) ) \  , \  \mathcal{P} ( \mathcal{M}^{C]} ) \  \Bigr) \  \  f_{ABC}}\hspace{1cm} \nonumber \\
&=& \Bigl( \  \mathcal{P}^3 ( \mathcal{M}^{[A} ) \  , \  \mathcal{P}^2 ( \mathcal{M}^B ) \  , \  \mathcal{P} ( \mathcal{M}^{C]} ) \  \Bigr) \  \  f_{ABC} \\
&=& \Bigl( \  \mathcal{M}^{[A} \  , \  \mathcal{P}^2 ( \mathcal{M}^B ) \  , \  \mathcal{P} ( \mathcal{M}^{C]} ) \  \Bigr) \  \  f_{ABC} \\
&=& \Bigl( \  \mathcal{P}^2 ( \mathcal{M}^{[B} ) \  , \  \mathcal{P} ( \mathcal{M}^C ) \  , \  \mathcal{M}^{A]} \  \Bigr) \  \  f_{BCA} \\
&=& S \nonumber
\end{eqnarray}
The reason why the cycle mapping is important is that the invariance of the action under this mapping has a deep connection with the {\itshape supersymmetry}. Basically, we would like to think that the specific components of $\mathcal{M}^A \in \Vec{\mathfrak{J}^c}$ are divided as follows, 
\begin{eqnarray}
\mathcal{M}^A
\,=\,
\left(
      \begin{array}{ccc}
      \mathcal{A}_1{}^A & \Phi_3{}^A & \bar{\Phi}_2{}^A \\
      \bar{\Phi}_3{}^A & \mathcal{A}_2{}^A & \Phi_1{}^A \\
      \Phi_2{}^A & \bar{\Phi}_1{}^A & \mathcal{A}_3{}^A
      \end{array}
\right)
\,\,\equiv\,
\left(
      \begin{array}{cc}
      \mathcal{W}^A & \Psi^A \\
      \Psi^\ddagger{}^A & v^A
      \end{array}
\right) \  \  , \label{eq:susy}
\end{eqnarray}
where $\mathcal{W}^A$, $\Psi^A$, and $v^A$ are defined by 
\begin{eqnarray}
\mathcal{W}^A
=
\left(
      \begin{array}{cc}
      \mathcal{A}_1{}^A & \Phi_3{}^A \\
      \bar{\Phi}_3{}^A & \mathcal{A}_2{}^A
      \end{array}
\right)
\  , \quad \Psi^A
=
\left(
      \begin{array}{c}
      \bar{\Phi}_2{}^A \\
      \Phi_1{}^A
      \end{array}
\right)
\  , \quad v^A
=
\mathcal{A}_3{}^A \  .
\end{eqnarray}
We would like to consider that $\mathcal{W}^A$ and $v^A$ are bosonic fields, and $\Psi^A$ are fermionic fields in the long run. Unfortunately, however, $\Psi^A$ are still bosonic fields at this stage. This is a problem common to all theories which are based on the exceptional Jordan algebra $\Vec{\mathfrak{J}}$ or the complex exceptional Jordan algebra $\Vec{\mathfrak{J}^c}$. Of course, if one would like to avoid this problem, one may consider the theory based on the Jordan algebra $\Vec{\mathfrak{j}}$ or the complex Jordan algebra $\Vec{\mathfrak{j}^c}$ and may prepare fermions made up of Grassmann variables separately. However, the groups like $F_4$ or $E_6$ are very attractive, and the idea that one multiplet is pushed into a single matrix, which is an element of $\Vec{\mathfrak{J}}$ or $\Vec{\mathfrak{J}^c}$, is mathematically beautiful. Therefore we have to prepare a mechanism separately, which introduces an anticommuting factor into the theory. An example of the way is that we impose different boundary conditions when we compactify the theory. It is a future problem whether other mechanisms exist. This is very interesting, because there is a close resemblance between this situation and that of the fractional quantum hall effects of condensed matter systems. It is necessary to keep in mind that it is possible that our definition of spinors itself is still incomplete. For another, this situation may have a connection with `Bosonic M Theory' \cite{Horowitz:2001gn}. Anyway, the expression (\ref{eq:susy}) is the reason why we would like to relate the cycle mapping to the supersymmetry.
%
\section{Constraints of the model}

We next study the constraints of our model. If this algebraically defined theory has a deep connection with some geometry, it must be essentially the {\itshape non-associative geometry}. In particular, it is quite likely that this theory has its geometrical interpretation in the {\itshape projective geometry} because the group we consider here is the compact $E_6$ group. For example, the non-compact group $E_{6(-26)}$, which is also one of the groups of type $E_6$, is the projective transformation group of Graves-Cayley projective plane $\Vec{\mathfrak{C}} P_2$ \cite{Freudenthal:1951gf}. Of course, little is known about the geometrical interpretation of our model at this stage. However, in order to pursue the geometrical interpretation of the theory in due course, it is important to refer to the fact that this model is a constrained system first.

The reason why additional conditions are imposed on our model is that the group we consider here is especially the {\itshape compact} $E_6$ group of type $E_6$. The constraints result from the following condition.
\begin{eqnarray}
\langle \alpha X , \alpha Y \rangle &=& \langle X , Y \rangle \\
& & {} \qquad ( X,Y \in \Vec{\mathfrak{J}^c} ) \nonumber
\end{eqnarray}
Thanks to this condition, one invariant under the $E_6$ mapping is introduced into the theory.
\begin{eqnarray}
\left\{
    \begin{array}{l}
    invariant_{E_6} = \langle \mathcal{M}^A , \mathcal{M}^B \rangle \quad \in \quad \mathbf{C} \\
    \\
    \langle \mathcal{M}^B , \mathcal{M}^A \rangle = \langle \mathcal{M}^A , \mathcal{M}^B \rangle {}^\ast
    \end{array}
\right.
\end{eqnarray}
Clearly, we can couple another invariant, which is the $\delta$-term concerning $\Vec{\mathcal{G}}$, to this. If we consider $\Vec{\mathcal{G}} = u(N)$ for example, summing up indices leads us to the following expression.
\begin{eqnarray}
invariant_{E_6 \times U(N)} &=& 2 \  \langle \mathcal{M}^A , \mathcal{M}^B \rangle \  tr ( \mathbf{T}_A \mathbf{T}_B ) \\
&=& \langle \mathcal{M}^A , \mathcal{M}^A \rangle \quad \in \quad \mathbf{R} \\
&\ge& 0
\end{eqnarray}
The crucial point to observe here is that {\itshape this invariant satisfies a positivity condition}: \ in general $\langle \mathcal{M}^A , \mathcal{M}^A \rangle \ge 0$, and vanishes if and only if $\mathcal{M}^A = 0$. Although this invariant is the quantity defined on the product space $\Vec{\mathfrak{J}^c} \times \Vec{\mathcal{G}}$, the resulting structure is, in a sense, quite similar to that of the physical Hilbert space. Since there is a cycle mapping, this space is, as it were, the Hilbert space with the spinor structure. In passing, we can rewrite above quantity with emphasis on $U(N)$, 
\begin{eqnarray}
invariant_{E_6 \times U(N)} &=& \sum_{I=1}^{3} \bigl( \mathcal{A}_I^\ast{}^A{} \mathcal{A}_I{}^B + 2\langle \Phi_I{}^A , \Phi_I{}^B \rangle \bigr) \  tr ( \mathbf{T}_A \mathbf{T}_B ) \\
&=& tr_{N \times N} \Bigl( \  ( \mathcal{A}_I^\ast{}^A \mathcal{A}_I{}^B + 2 \Phi_{\tilde{i}I}^\ast{}^A \Phi_{\tilde{i}I}{}^B ) \  \mathbf{T}_A \mathbf{T}_B \  \Bigr) \\
&\ge& 0 \  ,
\end{eqnarray}
where $N \to \infty$, or finite but boundlessly large.

Furthermore, we can consider the following combinations too.
\begin{eqnarray}
\langle \alpha \  \mathcal{P}^2 ( \mathcal{M}^A ) , \alpha \  \mathcal{P} ( \mathcal{M}^B ) \rangle &=& \langle \mathcal{P}^2 ( \mathcal{M}^A ) , \mathcal{P} ( \mathcal{M}^B ) \rangle \\
\langle \alpha \  \mathcal{P} ( \mathcal{M}^A ) , \alpha \  \mathcal{M}^B \rangle &=& \langle \mathcal{P} ( \mathcal{M}^A ) , \mathcal{M}^B \rangle \\
\langle \alpha \  \mathcal{M}^A , \alpha \  \mathcal{P}^2 ( \mathcal{M}^B ) \rangle &=& \langle \mathcal{M}^A , \mathcal{P}^2 ( \mathcal{M}^B ) \rangle
\end{eqnarray}

Next, let us consider more general case. Two $\mathcal{M}$'s can be independent this time. In this case, the invariant is as follows.
\begin{eqnarray}
\left\{
    \begin{array}{l}
    invariant_{E_6} = \langle \mathcal{M}^A , \mathcal{M}\Vec{'}^B \rangle \quad \in \quad \mathbf{C} \\
    \\
    \langle \mathcal{M}\Vec{'}^B , \mathcal{M}^A \rangle = \langle \mathcal{M}^A , \mathcal{M}\Vec{'}^B \rangle {}^\ast
    \end{array}
\right.
\end{eqnarray}
Therefore we can couple the $\delta$-term concerning $\Vec{\mathcal{G}}$ to this quantity in the same way as previous case.
\begin{eqnarray}
\left\{
    \begin{array}{l}
    invariant_{E_6 \times U(N)} = \langle \mathcal{M}^A , \mathcal{M}\Vec{'}^A \rangle \quad \in \quad \mathbf{C} \\
    \\
    \langle \mathcal{M}\Vec{'}^A , \mathcal{M}^A \rangle = \langle \mathcal{M}^A , \mathcal{M}\Vec{'}^A \rangle {}^\ast
    \end{array}
\right. \label{eq:invariant}
\end{eqnarray}
This invariant is, in general, a complex number. The second expression of (\ref{eq:invariant}) indicates that $\langle \mathcal{M}\Vec{'}^A , \mathcal{M}^A \rangle$ and $\langle \mathcal{M}^A , \mathcal{M}\Vec{'}^A \rangle$ are complex conjugates of each other.

Of course, we may also consider the following combinations as before.
\begin{eqnarray}
\langle \alpha \  \mathcal{P}^2 ( \mathcal{M}^A ) , \alpha \  \mathcal{P} ( \mathcal{M}\Vec{'}^B ) \rangle &=& \langle \mathcal{P}^2 ( \mathcal{M}^A ) , \mathcal{P} ( \mathcal{M}\Vec{'}^B ) \rangle \\
\langle \alpha \  \mathcal{P} ( \mathcal{M}^A ) , \alpha \  \mathcal{M}\Vec{'}^B \rangle &=& \langle \mathcal{P} ( \mathcal{M}^A ) , \mathcal{M}\Vec{'}^B \rangle \\
\langle \alpha \  \mathcal{M}^A , \alpha \  \mathcal{P}^2 ( \mathcal{M}\Vec{'}^B ) \rangle &=& \langle \mathcal{M}^A , \mathcal{P}^2 ( \mathcal{M}\Vec{'}^B ) \rangle
\end{eqnarray}

The author does not know the geometrical interpretation of the simply connected compact exceptional Lie group $E_6$. The attempt to relate the model based on the {\itshape compact} group $E_6$ to some geometry is a very exciting theme. It is quite likely that the theory based on $E_6$ is closely connected with the topological theory, because although the action was originally constructed from the {\itshape algebraic} invariant on $E_6$ mapping (i.e. cubic form), it has a form which is similar to the Chern-Simons theory automatically. To take a hypothetical example, if we can give the Freudenthal multiplication $X \times Y$ a definite interpretation such as the {\itshape outer product} in the space, we might be able to give the action a geometrical interpretation such as some kind of {\itshape volume}. What seems to be lacking is the knowledge of the projective geometry or the non-associative geometry. Therefore, we get the feeling that we had better pay attention to the advancement of these fields.
%
\section{One of the classical solutions and \,$T^3$ compactification}

One way to study the dynamics of the theory is the compactification on directions. In this section, we follow Smolin's arguments. We investigate one of the classical solutions of our model, and compactify the theory on the three directions. As a result, we can have the same argument as Smolin \cite{Smolin:2001wc} which derives an effective action similar to the matrix string theory. The main difference is that our model has twice as many degrees of freedom as Smolin's model has because we are considering $E_6$ instead of $F_4$.

To begin with, we represent the matrix elements $(\Vec{\mathcal{A}}_I)^P{}_Q$, where $P$ stands for the index of `row' and $Q$ stands for the index of `column', of $N \times N$ square matrices $\Vec{\mathcal{A}}_I$ as $\mathcal{A}_I{}^P_Q$. Then, let us view $\Vec{\mathcal{G}}$ as a product space which is made up of four parts. Accordingly, we can give the one-to-one correspondence between $P,Q$ and $(p_1 p_2 p_3 \tilde{P}),(q_1 q_2 q_3 \tilde{Q})$,
\begin{eqnarray}
\mathcal{A}_I{}^P_Q &\equiv& \mathcal{A}_I{}^{p_1}_{q_1}{}^{p_2}_{q_2}{}^{p_3}_{q_3}{}^{\tilde{P}}_{\tilde{Q}} \  ,
\end{eqnarray}
where $(p_I,q_I = -L_I,\cdots,0,\cdots,L_I) \, {}_{(I=1,2,3)} \, , \, (\tilde{P},\tilde{Q} = 1,\cdots,M)$, \,so that 
\begin{eqnarray}
N = \bigl( \prod_{I=1}^{3} (2 L_I + 1) \bigr) \  M \  .
\end{eqnarray}

Next, let us focus on one of the classical solutions of our model, given by 
\begin{eqnarray}
\left\{
    \begin{array}{c}
    \mathcal{A}_I{}^{p_1}_{q_1}{}^{p_2}_{q_2}{}^{p_3}_{q_3}{}^{\tilde{P}}_{\tilde{Q}} = P_I{}^{p_1}_{q_1}{}^{p_2}_{q_2}{}^{p_3}_{q_3} \  \delta^{\tilde{P}}_{\tilde{Q}} \\
    P_I{}^{p_1}_{q_1}{}^{p_2}_{q_2}{}^{p_3}_{q_3} = p_I \delta^{p_1}_{q_1} \delta^{p_2}_{q_2} \delta^{p_3}_{q_3}
    \end{array}
\right.
\end{eqnarray}
with other fields vanishing. This solution satisfies the following relations.
\begin{eqnarray}
[ ( \Vec{P}_I \otimes \Vec{1}_{M \times M} ) , ( \Vec{P}_J \otimes \Vec{1}_{M \times M} ) ] &=& 0 
\end{eqnarray}

Now, we expand the theory around this classical solution,
\begin{eqnarray}
\mathcal{A}_I{}^{p_1}_{q_1}{}^{p_2}_{q_2}{}^{p_3}_{q_3}{}^{\tilde{P}}_{\tilde{Q}} &=& P_I{}^{p_1}_{q_1}{}^{p_2}_{q_2}{}^{p_3}_{q_3} \  \delta^{\tilde{P}}_{\tilde{Q}} \  + \  a_I{}^{p_1}_{q_1}{}^{p_2}_{q_2}{}^{p_3}_{q_3}{}^{\tilde{P}}_{\tilde{Q}} \  ,
\end{eqnarray}
and then consider the mapping into the space of functional by using the usual matrix compactification procedure based on the complex Fourier series expansion, 
\begin{eqnarray}
tr_{N \times N} \Bigl( F [ P , G ] \Bigr) &=& \frac{1}{T} \oint \! dt \  tr_{M \times M} \Bigl( F(t) ( -i \frac{\partial G(t)}{\partial t} ) \Bigr) \label{eq:T^1} \  .
\end{eqnarray}
At this time, while the fields $a_I{}^{p_1}_{q_1}{}^{p_2}_{q_2}{}^{p_3}_{q_3}{}^{\tilde{P}}_{\tilde{Q}}$ and $\Phi_{\tilde{i}3}{}^{p_1}_{q_1}{}^{p_2}_{q_2}{}^{p_3}_{q_3}{}^{\tilde{P}}_{\tilde{Q}}$ are compactified as bosonic fields, the fields $\Phi_{\tilde{i}\alpha}{}^{p_1}_{q_1}{}^{p_2}_{q_2}{}^{p_3}_{q_3}{}^{\tilde{P}}_{\tilde{Q}} \  {}_{(\alpha = 1,2)}$ are compactified as fermionic fields.
\begin{eqnarray}
a_I{}^{(p_1+(2L_1+1))}_{(q_1+(2L_1+1))}{}^{p_2}_{q_2}{}^{p_3}_{q_3}{}^{\tilde{P}}_{\tilde{Q}} &=& + a_I{}^{p_1}_{q_1}{}^{p_2}_{q_2}{}^{p_3}_{q_3}{}^{\tilde{P}}_{\tilde{Q}} \\
\Phi_{\tilde{i}3}{}^{(p_1+(2L_1+1))}_{(q_1+(2L_1+1))}{}^{p_2}_{q_2}{}^{p_3}_{q_3}{}^{\tilde{P}}_{\tilde{Q}} &=& + \Phi_{\tilde{i}3}{}^{p_1}_{q_1}{}^{p_2}_{q_2}{}^{p_3}_{q_3}{}^{\tilde{P}}_{\tilde{Q}} \\
\Phi_{\tilde{i}\alpha}{}^{(p_1+(2L_1+1))}_{(q_1+(2L_1+1))}{}^{p_2}_{q_2}{}^{p_3}_{q_3}{}^{\tilde{P}}_{\tilde{Q}} &=& - \Phi_{\tilde{i}\alpha}{}^{p_1}_{q_1}{}^{p_2}_{q_2}{}^{p_3}_{q_3}{}^{\tilde{P}}_{\tilde{Q}}
\end{eqnarray}
The $x^2$ and $x^3$ directions are also compactified with the same signs.

Under these conditions, the action of the theory becomes 
\begin{eqnarray}
S &=& - \frac{3}{2 (T_1 T_2 T_3)} \  \oint_{T^3} \! d^3x \  tr_{M \times M} \Bigl( \  \  \epsilon^{IJK} \  ( \  \Vec{a}_I \partial_J \Vec{a}_K + \frac{2i}{3} \Vec{a}_I \Vec{a}_J \Vec{a}_K \  ) \nonumber \\
& & {} \qquad\quad\quad + 2 \  ( \  - \Vec{\Phi}_{\tilde{i}1} [ \mathcal{D}_3 , \Vec{\Phi}_{\tilde{i}2} ] + \Vec{\Phi}_{\tilde{i}2} [ \mathcal{D}_3 , \Vec{\Phi}_{\tilde{i}1} ] \  ) \nonumber \\
& & {} \qquad\quad\quad -4i \  ( \  \Vec{\Phi}_{01} [ \Vec{\Phi}_{03} , \Vec{\Phi}_{02} ] - \Vec{\Phi}_{i1} [ \Vec{\Phi}_{03} , \Vec{\Phi}_{i2} ] - \Vec{\Phi}_{i1} [ \Vec{\Phi}_{i3} , \Vec{\Phi}_{02} ] - \Vec{\Phi}_{01} [ \Vec{\Phi}_{i3} , \Vec{\Phi}_{i2} ] \nonumber \\
& & {} \qquad\qquad\qquad\quad - \sigma_{ijk} \Vec{\Phi}_{i1} [ \Vec{\Phi}_{j3} , \Vec{\Phi}_{k2} ] + \frac{1}{3} \sigma_{ijk} \Vec{\Phi}_{i3} [ \Vec{\Phi}_{j3} , \Vec{\Phi}_{k3} ] \  ) \  \  \Bigr) \label{eq:T^3-1} \\
&=& - \frac{3}{2 \Lambda} \  \oint_{T^3} \! d^3x \  tr_{M \times M} \Bigl( \  \  \epsilon^{IJK} \  ( \  \Vec{a}_I \partial_J \Vec{a}_K + \frac{2i}{3} \Vec{a}_I \Vec{a}_J \Vec{a}_K \  ) \nonumber \\
& & {} \qquad\qquad\qquad\qquad\qquad - 2 \  \epsilon^{\alpha \beta} \  \Vec{\Phi}_{\tilde{i}\alpha} [ \mathcal{D}_3 , \Vec{\Phi}_{\tilde{i}\beta} ] - \frac{8i}{3} \sigma_{ijk} \Vec{\Phi}_{i3} \Vec{\Phi}_{j3} \Vec{\Phi}_{k3} \  \  \Bigr) \nonumber \\
& & {} \qquad\qquad\qquad - 4i \  \Vec{Re^c} ( \Phi_3{}^{\tilde{P}}_{\tilde{Q}} \Phi_2{}^{\tilde{Q}}_{\tilde{R}} \Phi_1{}^{\tilde{R}}_{\tilde{P}} ) + 4i \  \Vec{Re^c} ( \Phi_1{}^{\tilde{P}}_{\tilde{Q}} \Phi_2{}^{\tilde{Q}}_{\tilde{R}} \Phi_3{}^{\tilde{R}}_{\tilde{P}} ) \label{eq:T^3-2}
\end{eqnarray}
in the $L_I \to \infty$ limits, where $T_I = l^{(I)} (2L_I+1)$, with $T_I$ held fixed and $l^{(I)} \to 0$. The dimensional scales $l^{(I)}$ are introduced in order to adjust the physical dimensions. The point to observe here is that the coefficient $\frac{1}{(T_1 T_2 T_3)}$ resulting from the compactification on directions fills the role of a cosmological term $\frac{1}{\Lambda}$. 

Incidentally, the expressions in terms of continuous functions like (\ref{eq:T^1}), (\ref{eq:T^3-1}) and (\ref{eq:T^3-2}) are symbolical notations. It would be rather essentially proper to think that there exists an $l_{Planck}$, which is sufficiently small but finite, such that $l^{(I)} \ge l_{Planck}$. In brief, this is to say that a minimum length is introduced into the space-time itself. Accordingly $L_I$ are also very large but finite. In consequence, such concept as {\itshape universality class} seems to be unacceptable because we cannot achieve the genuine continuum limit. Applying the concept of universality class to the matrix model implies that the matrix model drops its position down to a mere regularization of the continuum theory. We do not take the position that we consider the matrix model to be the regularization of the continuum theory. We take a view that it is the expression in terms of matrices that is rather proper descriptive language. The reason why Wilson's method achieved a great success is that the field theory itself was a low-energy effective theory. Of course it is possible that such concepts as {\itshape nonstandard number} and {\itshape nonstandard analysis} are not irrelevant to the future physics. However, what needs to be emphasized here is that we must not handle the micro-world as an object of `regulation' in order to suit macro-phenomena to our own convenience. Clearly, the macro-world is in existence as a consequence of phenomena of the micro-world. The research worker of the elementary particle should pursue the structure of minute world to the bitter end, and now, it would be wise for us to abandon the concept of the space-time {\itshape continuum} as a product of illusion.

Let us now return to our main concern. One process leading to (\ref{eq:T^3-1}) is that we eliminate the terms which comprise the odd degree with respect to the fermionic fields $\Phi_{\tilde{i}\alpha}{}^{p_1}_{q_1}{}^{p_2}_{q_2}{}^{p_3}_{q_3}{}^{\tilde{P}}_{\tilde{Q}} \  {}_{(\alpha = 1,2)}$ from the compactified action because they contradict the boundary conditions. It is now easy to show that what is true for Smolin's model is true for our model as well. (See Ref.\cite{Smolin:2001wc}, Section 5.) \ If we take the limit $T_3 \to T_{Planck}$ and then drop terms in $\partial_3$ and $a_3$ intentionally, aside from the coefficient of each term we can expect that from analyses of symmetries and power counting the resulting 2-dimensional effective action results in what is similar to the action of the matrix string theory at the one loop level.
\begin{eqnarray}
I_{eff} &=& \oint_{T^2} \! d^2x \  tr_{M \times M} \Bigl( \  \Vec{\Phi}_{\tilde{i}\alpha} \sigma^{\mu \alpha \beta}{}_{\tilde{i} \tilde{j}} [ \mathcal{D}_{\mu} , \Vec{\Phi}_{\tilde{j}\beta} ] + \Vec{\Phi}_{\tilde{i}1} \gamma^{\tilde{k}}{}_{\tilde{i} \tilde{j}} [ \Vec{\Phi}_{\tilde{k}3} , \Vec{\Phi}_{\tilde{j}2} ] + \sigma_{ijk} \Vec{\Phi}_{i3} \Vec{\Phi}_{j3} \Vec{\Phi}_{k3} \nonumber \\
& & {} \qquad\qquad\qquad\quad\  + ( \Vec{f}_{\mu \nu} )^2 + ( \mathcal{D}_{\mu} \Vec{\Phi}_{\tilde{i}3} )^2 + \rho_{ijkl} [ \Vec{\Phi}_{i3} , \Vec{\Phi}_{j3} ] [ \Vec{\Phi}_{k3} , \Vec{\Phi}_{l3} ] \  \Bigr)
\end{eqnarray}
One of the differences between this action and that of the matrix string theory is the form of the four-matrix interaction terms as has been pointed out by Smolin. This is, so to speak, a matrix string-like theory based on $G_2{}\Vec{^c}$. The other crucial point is that our model has twice as many degrees of freedom as Smolin's model has because we are considering $E_6$ instead of $F_4$. This trouble always follows us as long as we handle $E_6$. We will take the argument about this matter up some other time.

Lastly, the author would like to mention the possibility that the 3-dimensional compactified action (\ref{eq:T^3-2}) might be associated with not only the constructive formulation of the string theory but also the loop quantum gravity which is another hopeful approach to the unified theory. Let us now exponentiate the action (\ref{eq:T^3-2}) and suppose the following quantity.
\begin{eqnarray}
\Vec{\mit \Psi}_\Lambda[a,\Phi] &\equiv& e^{iS} \\
&=& e^{-i \frac{3}{2 \Lambda} \bigl( I_{cs}(a) + I(\Phi) \bigr)} \\
I_{cs}(a) &=& \oint_{T^3} \! d^3x \  \epsilon^{IJK} \  tr_{M \times M} \Bigl( \  \Vec{a}_I \partial_J \Vec{a}_K + \frac{2i}{3} \Vec{a}_I \Vec{a}_J \Vec{a}_K \  \Bigr)
\end{eqnarray}
Except for the facts that this quantity contains the fermionic terms and $U(M)$ takes the place of $SU(2)$, it is similar to Kodama's wave function \cite{Kodama:1988yf,Kodama:1990sc,Smolin:1995qb} which is an exact solution to all the constraints of the loop quantum gravity in the case of non-zero cosmological constant. In fact, the overall factor is $-i \frac{3}{2 \Lambda}$ which is the same as Kodama's wave function. We did not use any approximation, nevertheless the same factor is directly derived from our model. Therefore, it seems quite probable that there exists something like an $U(M)$ {\itshape generalized} loop quantum gravity whose physical state is $\Vec{\mit \Psi}_\Lambda[a,\Phi]$.
%
%
%
%
%
%
\section{Conclusion and Discussion}

In this short paper, we have introduced a new matrix model based on the simply connected compact exceptional Lie group $E_6$, and discovered that the Chern-Simons like term is also derived from the {\itshape cubic form}. This theory modeled itself on Smolin's approach based on the groups of type $F_4$ \cite{Smolin:2001wc}. We have adopted the {\itshape cubic form} in place of the {\itshape trilinear form}. The {\itshape cubic form} is a quite different cubic linear form from the {\itshape trilinear form}. Our model has twice as many degrees of freedom as Smolin's model has because we consider {\itshape compact} $E_6$ instead of $F_4$. That point aside, we can have the same argument as Smolin which derives an effective action similar to the matrix string theory. An important point to emphasize is that it is characteristic of $E_6$ and $F_4$ to derive the Chern-Simons type action using this method, because $G_2$, $E_7$ and $E_8$ have no cubic linear forms which are made up of the pure exceptional Jordan algebra alone just like the {\itshape trilinear form} on $F_4$ or the {\itshape cubic form} on $E_6$. One way to introduce the cosmological term or the coupling constant into the theory is the compactification on directions. Of course, what we have reported here is just the first step in the analysis of the model, and many things need to be investigated. However, as we have seen, this model has several very interesting characteristics. Therefore it is quite likely that this theory will evolve in the future.

The problem to be specially considered is the quantization. It may be difficult to path-integrate as usual because the action of the theory is an essentially complex action. Consequently, it seems that it is necessary to reexamine the {\itshape canonical formalism} such as the loop quantum gravity. The author is very interested in the relation between the exponentiated quantity of the action (\ref{eq:E_6}) and some kind of quantity like a bi-local expression of Dirac equation. If one point is decided as a fiducial point, another point might be indicated using twistor-like method.

Besides, because this model is written in `cubic' terms with respect to the fields, there cannot exist the term such as $R_{\mu\nu}{}^{\mu\nu}$ essentially. In consequence, we are inevitably obliged to take the position of the {\itshape induced gravity}. To put it the other way round, however, thanks to being cubic, the theory becomes what is close to the topological theory; and that is also one of the interesting properties of this model. Namely, there is a strong possibility that this theory is defined as a background independent theory from the very beginning. Therefore it is quite likely that other matrix models such as the BFSS model and the IKKT model are reproduced via expansions around specific backgrounds of this model.

Moreover, it seems quite probable that this algebraically defined model has a geometrical interpretation. The existence of the projective geometry can be seen off and on behind the Freudenthal multiplication or the cubic form. Although, depending on how things go, we might have to introduce even the Freudenthal manifold $\mathfrak{M}$ which is needed to understand the groups of type $E_7$ and type $E_8$, the attempt to relate this theory to some geometry is one of the most exciting subjects.

Furthermore, as discussed in section 5, the author would like to emphasize that there is considerable validity in considering the physical Hilbert space to be a {\itshape product space} composed of two parts $\Vec{\mathfrak{J}^c}$ and $\Vec{\mathcal{G}}$. Because there exists a cycle mapping, the resulting product space has a structure such that the concept of the spinor is introduced into the infinitely dimensional Hilbert space itself. The author does not believe it is a coincidence. It is likely that $\Vec{\mathfrak{J}^c}$ describes some degrees of freedom belonging to some internal structure in the each point of the space, and $\Vec{\mathcal{G}}$ plays a role of the network to the each point of the space. What is called {\itshape the postulate of positive definite metric} is, from the viewpoint of our model, merely an immediate consequence of the fact that our universe is {\itshape compact}.

By the way, we have considered thus far only the combination of the cubic form and the structure constant as an action of the $E_6$ matrix model. From the viewpoint of the construction of invariants, however, we can take not only $f$-coupling but also $d$-coupling into account, 
\begin{eqnarray}
S &=& \alpha \  \Bigl( \  \mathcal{P}^2 ( \mathcal{M}^{[A} ) \  , \  \mathcal{P} ( \mathcal{M}^B ) \  , \  \mathcal{M}^{C]} \  \Bigr) \  f_{ABC} \nonumber \\
  & & {} + \beta \  \Bigl( \  \mathcal{P}^2 ( \mathcal{M}^{(A} ) \  , \  \mathcal{P} ( \mathcal{M}^B ) \  , \  \mathcal{M}^{C)} \  \Bigr) \  d_{ABC} \label{eq:E_6fd}
\end{eqnarray}
where $( \cdots )$ denotes the {\itshape weight}-1 symmetrization on indices, and $d_{ABC}$ is defined as follows.
\begin{eqnarray}
\{ \mathbf{T}_A , \mathbf{T}_B \} &=& d_{ABC} \  \mathbf{T}_C \\
d_{ABC} &=& 2 \  tr ( \mathbf{T}_A \{ \mathbf{T}_B , \mathbf{T}_C \} )
\end{eqnarray}
Although it is not yet clear whether considering such an action is invaluable, it is possible that some interesting physics concerned with anomalies are gained because the third-rank symmetric tensor is introduced into the theory. If some kinds of anomalies are given rise to in our model, the cause of those seems to be $\Vec{\mathcal{G}}$. Because $E_6$ itself is a safe group.

Incidentally, the theories expressed by such actions as (\ref{eq:E_6}) and (\ref{eq:E_6fd}) are the {\itshape global} $E_6$ matrix models. Consequently, we naturally hit upon an idea that we might be able to localize this symmetry. What is called the {\itshape local} $E_6$ matrix model is the theory which has an invariant action under the mixed transformation on $\mathfrak{e}_6$ and $\Vec{\mathcal{G}}$. To use Smolin's words, we can call this type of matrix model the `gauged' matrix model. In fact, if we localize Smolin's $osp(1|32;\mathbf{R})$ matrix model, the resulting theory is just the $u(1|16,16)$ matrix model. This $u(1|16,16)$ matrix model is a very beautiful model. Therefore, we get the feeling that we would like to carry out the same thing as this to the global $E_6$ matrix model. However, the attempt to construct the local $E_6$ matrix model has not been succeeded so far because of some mathematical difficulties.

Finally, there is a further question which needs to be asked: \ Our model has twice as many degrees of freedom as Smolin's model has. This trouble always follows us as long as we handle $E_6$, and it is a serious problem which cannot be avoided. This problem will be discussed on another occasion.
%
%
%
%
\subsection*{Acknowledgments} 
I would like to thank Ichiro Yokota for helpful comments. I am also grateful to Albert Schwarz for sending a note \cite{Kim:2001up} to me. Lee Smolin's talk presented at The 10th Tohwa University International Symposium (July 3-7, 2001, Fukuoka, Japan) was my motive for starting this work.
%
%
%
%
\section*{Appendices}         
\appendix                                                  %
\section{Convention}

In this paper, repeated indices are generally summed unless otherwise indicated.

\subsection{(Anti-)symmetrization}
$[ \cdots ]$ denotes the {\itshape weight}-1 anti-symmetrization and \,$( \cdots )$ denotes the {\itshape weight}-1 symmetrization on indices as follows.
\begin{eqnarray}
X^{[A} Y^B Z^{C]}
&=&
\frac{1}{6} ( X^{A} Y^{B} Z^{C} + X^{B} Y^{C} Z^{A} + X^{C} Y^{A} Z^{B} \nonumber \\
& & {} - X^{C} Y^{B} Z^{A} - X^{B} Y^{A} Z^{C} - X^{A} Y^{C} Z^{B} ) \\
X^{(A} Y^B Z^{C)}
&=&
\frac{1}{6} ( X^{A} Y^{B} Z^{C} + X^{B} Y^{C} Z^{A} + X^{C} Y^{A} Z^{B} \nonumber \\
& & {} + X^{C} Y^{B} Z^{A} + X^{B} Y^{A} Z^{C} + X^{A} Y^{C} Z^{B} )
\end{eqnarray}
Therefore, contracting indices with the totally anti-symmetric tensor $A_{ABC}$ or the totally symmetric tensor $S_{ABC}$ results in the following ordinary summation.
\begin{eqnarray}
X^{[A} Y^B Z^{C]} \  A_{ABC}
&=& X^A Y^B Z^C \  A_{ABC} \\
X^{(A} Y^B Z^{C)} \  S_{ABC}
&=& X^A Y^B Z^C \  S_{ABC}
\end{eqnarray}

\subsection{Levi-Civita tensor in 2 dimensions}
The 2-dimensional Levi-Civita tensor is defined by 
\begin{eqnarray}
(\epsilon^{\alpha \beta}) = (\epsilon_{\alpha \beta}) = 
\left(
    \begin{array}{cc}
    0 & 1 \\
    -1 & 0
    \end{array}
\right) \  .
\end{eqnarray}

\subsection{Fundamental representation of $su(2)$}
The matrices which generate $su(2)$ are given by 
\begin{eqnarray}
\Vec{\tau}^1
&=& \frac{1}{2} \Vec{\sigma}^1
\  = \  \frac{1}{2}
        \left(
              \begin{array}{cc}
              0 & 1 \\
              1 & 0
              \end{array}
        \right) \\
\Vec{\tau}^2
&=& \frac{1}{2} \Vec{\sigma}^2
\  = \  \frac{1}{2}
        \left(
              \begin{array}{cc}
              0 & -i \\
              i & 0
              \end{array}
        \right) \\
\Vec{\tau}^3
&=& \frac{1}{2} \Vec{\sigma}^3
\  = \  \frac{1}{2}
        \left(
              \begin{array}{cc}
              1 & 0 \\
              0 & -1
              \end{array}
        \right) \  .
\end{eqnarray}
These satisfy 
\begin{eqnarray}
[ \Vec{\tau}^I , \Vec{\tau}^J ] &=& i \epsilon^{IJK} \  \Vec{\tau}^K \\
tr ( \Vec{\tau}^I \Vec{\tau}^J ) &=& \frac{1}{2} \delta^{IJ} \\
\epsilon^{IJK} &=& -2i \  tr ( \Vec{\tau}^I [ \Vec{\tau}^J , \Vec{\tau}^K ] ) \  .
\end{eqnarray}
Therefore, the 3-dimensional Levi-Civita tensor $\epsilon^{IJK} \  (I=1,2,3)$ is defined with 
\begin{eqnarray}
\epsilon^{123} = \epsilon_{123} = +1 \  ,
\end{eqnarray}
in which the exchange of the position between the upper suffix and the downstairs suffix does not make any sense.

\subsection{Fundamental representation of the Lie algebra $\Vec{\mathcal{G}}$}
The matrices which generate $\Vec{\mathcal{G}}$ satisfy 
\begin{eqnarray}
[ \mathbf{T}_A , \mathbf{T}_B ] &=& i f_{ABC} \  \mathbf{T}_C \\
tr ( \mathbf{T}_A \mathbf{T}_B ) &=& \frac{1}{2} \delta_{AB} \\
f_{ABC} &=& -2i \  tr ( \mathbf{T}_A [ \mathbf{T}_B , \mathbf{T}_C ] ) \  .
\end{eqnarray}
%
\section{Complex Graves-Cayley algebra $\Vec{\mathfrak{C}^c}$}

\subsection{Graves-Cayley algebra $\Vec{\mathfrak{C}}$}
\begin{figure}[htbp]                                         %
\begin{center}                                               %
\includegraphics[]{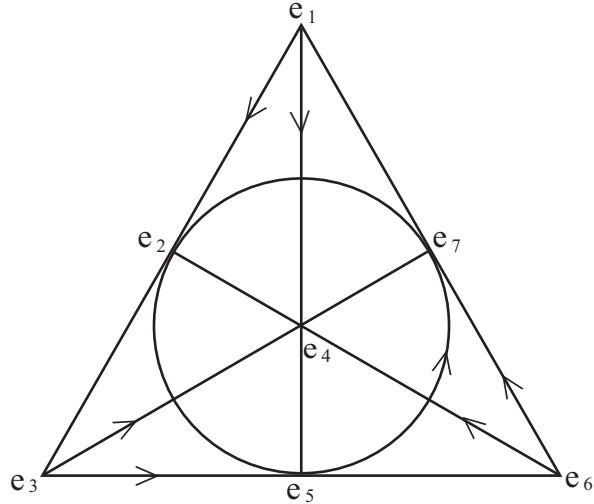}                             %
\end{center}                                                 %
\caption{Multiplication diagram for octonion}                %
\label{fig:octonion.eps}                                     %
\end{figure}

Let $\Vec{\mathfrak{C}} = \sum_{\tilde{i}=0}^{7} \, \mathbf{R} \, e_{\tilde{i}}$ be the Graves-Cayley algebra: \ $\Vec{\mathfrak{C}}$ is an 8-dimensional $\mathbf{R}$-vector space with the multiplication such that $e_0 = 1$ is the unit, \  $e_i{}^2 = -1 \  \  {}_{( i = 1,\cdots,7 )}$, \  $e_i e_j = - e_j e_i \  \  {}_{( 1 \le i \neq j \le 7 )}$ \  and $e_1 e_2 = e_3$, $e_1 e_4 = e_5$, $e_2 e_5 = e_7$, {\itshape etc}.\,. The element `$a$' of $\Vec{\mathfrak{C}}$ is called {\itshape octonion} (or {\itshape Graves-Cayley number}). The multiplication rule among the bases of $\Vec{\mathfrak{C}}$ can be represented in a diagram as Figure\,\ref{fig:octonion.eps}.
So if we take $e_1,e_2,e_3$ for example, 
\begin{equation}
    \begin{array}{c}
    e_1e_2=e_3 \  , \  e_2e_3=e_1 \  , \  e_3e_1=e_2 \\
    e_3e_2=-e_1 \  , \  e_2e_1=-e_3 \  , \  e_1e_3=-e_2 \  .
    \end{array}
\end{equation}
The same things can be said of other six lines. What has to be noticed is that this algebra is {\itshape non-associative} as well as non-commutative. It is often very useful to introduce the following notations, 
\begin{eqnarray}
e_ie_j &=& - \delta_{ij} + \sum_{k=1}^{7} \sigma_{ijk} e_k \  , \\
& &
\qquad\qquad ( i,j,k = 1,\cdots,7 ) \nonumber
\end{eqnarray}
where the $\sigma_{ijk}$ are totally anti-symmetric in indices, with values 1, 0, -1\,.\\
For instance, \,$\sigma_{ijk} = +1$ \,for \,$ijk = 123, \  356, \  671, \  145, \  347, \  642, \  257$\,.

In $\Vec{\mathfrak{C}}$, \,the conjugate $\bar{a}$ \,and the real part $\Vec{Re} ( a )$ \,are defined respectively as follows.
\begin{eqnarray}
a &\equiv& a_0 + \sum_{i=1}^{7} a_i e_i \\
\vspace{1cm} \nonumber \\
\bar{a} &=& \overline{a_0 + \sum_{i=1}^{7} a_i e_i} \\
&\equiv& a_0 - \sum_{i=1}^{7} a_i e_i \\
\vspace{1cm} \nonumber \\
\Vec{Re} ( a ) &\equiv& \frac{1}{2} ( a + \bar{a} ) \quad\quad \in \  \mathbf{R} \\
&=& a_0 \\
&=& \Vec{Re} ( \bar{a} )
\end{eqnarray}

Moreover, \,the inner product $( a , b ) \  \  ( a,b \in \Vec{\mathfrak{C}} )$ \,is defined by 
\begin{eqnarray}
( a , b ) &\equiv& a_0 b_0 + \sum_{i=1}^{7} a_i b_i \quad \in \  \mathbf{R} \\
&=& ( b , a ) \  .
\end{eqnarray}
Therefore we have 
\begin{eqnarray}
( a , a ) &=& (a_0)^2 + \sum_{i=1}^{7} (a_i)^2 \quad \in \  \mathbf{R} \\
&\ge& 0 \  .
\end{eqnarray}

\subsection{$\mathbf{C}$ and $\mathbf{H}$ \  in $\Vec{\mathfrak{C}}$}
\subsubsection{Complex number field in $\Vec{\mathfrak{C}}$}
The Graves-Cayley algebra $\Vec{\mathfrak{C}}$ contains the field of complex numbers $\mathbf{C}$.
\begin{eqnarray}
\mathbf{C} &=& \{ r_0 + r_4 e_4 \  | \  r_0,r_4 \in \mathbf{R} \}
\end{eqnarray}
\begin{eqnarray}
a &=& a_0 + \sum_{i=1}^{7} a_i e_i \\
&=& (a_0 + a_4 e_4) + (a_1 - a_5 e_4) e_1 + (a_2 + a_6 e_4) e_2 + (a_3 - a_7 e_4) e_3 \\
&=& c_0 + c_1 e_1 + c_2 e_2 + c_3 e_3 \\
& &
\quad\quad\quad\quad\quad c_k \in \mathbf{C} \quad\quad (k = 0,1,2,3) \nonumber
\end{eqnarray}

It must be noted that these complex numbers which have an imaginary unit `$e_4$' \,are independent of what are introduced in the following subsection whose imaginary unit is \,`$i$'\,.

\subsubsection{Quaternion field in $\Vec{\mathfrak{C}}$}
Furthermore, the Graves-Cayley algebra $\Vec{\mathfrak{C}}$ contains the field of quaternions $\mathbf{H}$ as well.
\begin{eqnarray}
\mathbf{H} &=& \{ r_0 + r_1 e_1 + r_2 e_2 + r_3 e_3 \  | \  r_0,r_1,r_2,r_3 \in \mathbf{R} \}
\end{eqnarray}
\begin{eqnarray}
a &=& a_0 + \sum_{i=1}^{7} a_i e_i \\
&=& (a_0 + a_1 e_1 + a_2 e_2 + a_3 e_3) + (a_4 + a_5 e_1 - a_6 e_2 + a_7 e_3) e_4 \\
&=& q_0 + q_4 e_4 \\
& &
\quad\quad\quad\quad q_k \in \mathbf{H} \quad\quad (k = 0,4) \nonumber
\end{eqnarray}

\subsection{Complex Graves-Cayley algebra $\Vec{\mathfrak{C}^c}$}
Let $\Vec{\mathfrak{C}^c}$, called the complex Graves-Cayley algebra, be the complexification of $\Vec{\mathfrak{C}}$.
\begin{eqnarray}
\Vec{\mathfrak{C}^c} = \{ a + ib \  | \  a,b \in \Vec{\mathfrak{C}} \ , \  i^2=-1 \}
\end{eqnarray}

Here, we should notice that \,`$i$' \,is introduced as a different imaginary unit from `$e_4$' which is that of the complex number field embedded in $\Vec{\mathfrak{C}}$ as mentioned in the previous subsection. This `$i$' commutes to all the $e_{\tilde{i}}$'s ${}_{( \tilde{i} = 0,\cdots,7 )}$\,.

In $\Vec{\mathfrak{C}^c}$, \,the conjugate $\bar{x}$ \,and the real part $\Vec{Re^c} ( x )$ \,are defined respectively as follows.
\begin{eqnarray}
x &=& (a_0 + \sum_{i=1}^{7} a_i e_i) + i(b_0 + \sum_{i=1}^{7} b_i e_i) \nonumber \\
&=& (a_0 + ib_0) + \sum_{i=1}^{7} (a_i + ib_i) e_i \nonumber \\
&\equiv& x_0 + \sum_{i=1}^{7} x_i e_i \\
\vspace{1cm} \nonumber \\
\bar{x} &=& \overline{(a_0 + \sum_{i=1}^{7} a_i e_i)} + i \overline{(b_0 + \sum_{i=1}^{7} b_i e_i)} \nonumber \\
&=& (a_0 + ib_0) - \sum_{i=1}^{7} (a_i + ib_i) e_i \nonumber \\
&\equiv& x_0 - \sum_{i=1}^{7} x_i e_i \\
\vspace{1cm} \nonumber \\
\Vec{Re^c} ( x ) &\equiv& \frac{1}{2} ( x + \bar{x} ) \quad\quad \in \  \mathbf{C} \\
&=& x_0 \\
&=& a_0 + ib_0 \\
&=& \Vec{Re^c} ( \bar{x} )
\end{eqnarray}

Moreover, \,for any two elements $x = a + ib$ and $y = c + id$ \,of $\Vec{\mathfrak{C}^c}$, \,the inner product $( x , y )$ \,is defined by 
\begin{eqnarray}
( x , y ) &\equiv& x_0 y_0 + \sum_{i=1}^{7} x_i y_i \quad \in \  \mathbf{C} \\
&=& (a_0 + ib_0) (c_0 + id_0) + \sum_{i=1}^{7} (a_i + ib_i) (c_i + id_i) \\
&=& ( y , x ) \  .
\end{eqnarray}
Therefore we have 
\begin{eqnarray}
( x , x ) &=& (x_0)^2 + \sum_{i=1}^{7} (x_i)^2 \quad \in \  \mathbf{C} \\
&=& \bar{x} x \  = \  x \bar{x} \  .
\end{eqnarray}

Furthermore, \,in $\Vec{\mathfrak{C}^c}$, \,the hermitian product $\langle x , y \rangle \  \  ( x,y \in \Vec{\mathfrak{C}^c} )$ \,is defined by 
\begin{eqnarray}
\langle x , y \rangle &\equiv& ( x^\ast , y ) \quad \in \  \mathbf{C} \\
&=& (a_0 - ib_0) (c_0 + id_0) + \sum_{i=1}^{7} (a_i - ib_i) (c_i + id_i) \  ,
\end{eqnarray}
where $(\cdots)^\ast$, \,called the complex conjugation with respect to `$i$', \,is defined by the following mapping.
\begin{eqnarray}
( a + ib )^\ast &\equiv& a - ib \\
& &
\quad\quad\quad\quad a,b \in \Vec{\mathfrak{C}} \nonumber
\end{eqnarray}
Therefore we have 
\begin{eqnarray}
x^\ast &=& x_0^\ast + \sum_{i=1}^{7} x_i^\ast e_i \  .
\end{eqnarray}
Naturally, \,we must not confuse this complex conjugation $(\cdots)^\ast$ \,with the octonionic conjugation $\overline{(\cdots)}$\,. \,An example is 
\begin{eqnarray}
\Vec{Re^c} ( x ) = \Vec{Re^c} ( \bar{x} ) \neq \Vec{Re^c} ( x^\ast ) \  .
\end{eqnarray}
Consequently, \,for any element $x = a + ib$ \,of $\Vec{\mathfrak{C}^c}$, \,we have 
\begin{eqnarray}
\langle x , x \rangle &=& \Bigl( (a_0)^2 + \sum_{i=1}^{7} (a_i)^2 \Bigr) + \Bigl( (b_0)^2 + \sum_{i=1}^{7} (b_i)^2 \Bigr) \\
\vspace{1cm} \nonumber \\
&=& ( a , a ) + ( b , b ) \quad \in \  \mathbf{R} \\
&\ge& 0 \  .
\end{eqnarray}

\subsection{Some helpful formulas on elements of $\Vec{\mathfrak{C}^c}$}
We can use the following formulas for any $w,x,y,z \in \Vec{\mathfrak{C}^c}$.
\begin{eqnarray}
(x^\ast)^\ast &=& x \\
( x + y )^\ast &=& x^\ast + y^\ast \\
(xy)^\ast &=& x^\ast y^\ast \\
\overline{(\bar{x})} &=& x \\
\overline{( x + y )} &=& \bar{x} + \bar{y} \\
\overline{(xy)} &=& \bar{y} \bar{x} \\
( x , y ) &=& \frac{1}{2} ( \bar{x} y + \bar{y} x ) \  = \  \frac{1}{2} ( x \bar{y} + y \bar{x} ) \\
&=& \Vec{Re^c} ( \bar{x} y ) \  = \  \Vec{Re^c} ( x \bar{y} ) \\
( x , y z ) &=& ( y , x \bar{z} ) \  = \  ( z , \bar{y} x ) \\
( x , y ) z &=& \frac{1}{2} \{ \bar{x} ( y z ) + \bar{y} ( x z ) \} \  = \  \frac{1}{2} \{ ( z y ) \bar{x} + ( z x ) \bar{y} \} \\
( w , x )( y , z ) &=& \frac{1}{2} \{ ( w y , x z ) + ( x y , w z ) \} \  = \  \frac{1}{2} \{ ( y w , z x ) + ( y x , z w ) \} \\
\Vec{Re^c} ( x y ) &=& x_0 y_0 - x_i y_i \\
&=& \frac{1}{2} ( xy + \bar{y}\bar{x} ) \  = \  \frac{1}{2} ( \bar{x}\bar{y} + yx ) \\
&=& \Vec{Re^c} ( y x ) \\
\Vec{Re^c} ( x y z ) &\equiv& \Vec{Re^c} ( x (y z) ) \  = \  \Vec{Re^c} ( (x y) z ) \\
&=& x_0 y_0 z_0 - x_0 y_i z_i - x_i y_0 z_i - x_i y_i z_0 - x_i y_j z_k \sigma_{ijk} \\
&=& \frac{1}{2} ( x(yz) + (\bar{z}\bar{y})\bar{x} ) \  = \  \frac{1}{2} ( (xy)z + \bar{z}(\bar{y}\bar{x}) ) \\
&=& \Vec{Re^c} ( y z x ) \  = \  \Vec{Re^c} ( z x y ) \\
&=& \Vec{Re^c} ( z y x ) - 2 x_i y_j z_k \sigma_{ijk} \ \qquad\qquad (i,j,k = 1,\cdots,7) \\
( x y ) z &=& \Vec{Re^c} ( x y z ) \nonumber \\
& & {} + \Bigl( \  x_0 y_0 z_l + x_0 y_l z_0 + x_l y_0 z_0 - x_i y_i z_l \nonumber \\
& & {} + x_0 y_i z_j \sigma_{ijl} + x_i y_0 z_j \sigma_{ijl} + x_i y_j z_0 \sigma_{ijl} \nonumber \\
& & {} + x_i y_j z_k \sigma_{ijm} \sigma_{klm} \  \Bigr) e_l \\
x ( y z ) &=& \Vec{Re^c} ( x y z ) \nonumber \\
& & {} + \Bigl( \  x_0 y_0 z_l + x_0 y_l z_0 + x_l y_0 z_0 - x_l y_i z_i \nonumber \\
& & {} + x_0 y_i z_j \sigma_{ijl} + x_i y_0 z_j \sigma_{ijl} + x_i y_j z_0 \sigma_{ijl} \nonumber \\
& & {} - x_i y_j z_k \sigma_{jkm} \sigma_{ilm} \  \Bigr) e_l \\
( x y ) \bar{y} &=& x ( y \bar{y} ) \  = \  ( y \bar{y} ) x \  = \  y ( \bar{y} x ) \\
( x y ) \bar{x} &=& x ( y \bar{x} ) \  , \quad ( x y ) x \  = \  x ( y x ) \\
( x x ) y &=& x ( x y ) \  , \quad x ( y y ) \  = \  ( x y ) y \\
{ [x,y,z] } &\equiv& ( x y ) z - x ( y z ) \\
&=& x_i y_j z_k \  ( \sigma_{ijm} \sigma_{klm} + \sigma_{jkm} \sigma_{ilm} + \delta_{kj} \delta_{il} - \delta_{kl} \delta_{ij} ) \  e_l \\
&=& x_i y_j z_k \  ( \sigma_{ijm} \sigma_{klm} + \sigma_{jkm} \sigma_{ilm} + \sigma_{kim} \sigma_{jlm} ) \  e_l \\
&\equiv& x_i y_j z_k \  ( \rho_{ijkl} ) \  e_l \\
& &
\quad\qquad\qquad\qquad (\rho_{ijkl}: \ \mbox{completely antisymmetric}) \nonumber \\
{ [x,y,z] } &=& { [y,z,x] } \  = \  { [z,x,y] } \\
&=& { -[z,y,x] } \  = \  { -[y,x,z] } \  = \  { -[x,z,y] } \\
&=& { -[x,y,\bar{z}] } \  = \  { [\bar{z},\bar{y},\bar{x}] } \  = \  - \overline{ [x,y,z] } \\
\Vec{Re^c} ( \  { [x,y,z] } \  ) &=& 0 \\
( x y ) z + \overline{x ( y z )} &=& 2 \Vec{Re^c} ( x y z ) + { [x,y,z] } \\
\overline{( x y ) z} + x ( y z ) &=& 2 \Vec{Re^c} ( x y z ) - { [x,y,z] } \\
x ( y z ) + ( y z ) x &=& ( x y ) z + y ( z x ) \\
x ( y z ) + x ( z y ) &=& ( x y ) z + ( x z ) y \\
( x y ) z + ( y x ) z &=& x ( y z ) + y ( x z )
\end{eqnarray}
%
\section{Complex exceptional Jordan algebra $\Vec{\mathfrak{J}^c}$}

\subsection{Jordan algebra $\Vec{\mathfrak{j}}$}
We define $\Vec{\mathfrak{j}}$ \,as the Jordan algebra consisting of all $2 \times 2$ hermitian matrices $A$ \,with entries in the Graves-Cayley algebra $\Vec{\mathfrak{C}}$\,.
\begin{eqnarray}
\Vec{\mathfrak{j}} = \{ A \in M(2,\Vec{\mathfrak{C}}) \  | \  A^\ddagger = A \} \quad\quad \bigl( A^\ddagger \equiv (\bar{A})^T \bigr)
\end{eqnarray}
The specific components of $A$ \,can be written as follows.
\begin{eqnarray}
A
&=&
\left(
      \begin{array}{cc}
      Q_1 & \phi_3 \\
      \bar{\phi}_3 & Q_2 \\
      \end{array}
\right) \\
& &
\quad\quad\quad Q_I \in \mathbf{R} \ \quad \phi_3 \in \Vec{\mathfrak{C}} \quad\quad (I=1,2) \nonumber
\end{eqnarray}
Therefore, \,$\Vec{\mathfrak{j}}$ \,is a 10-dimensional $\mathbf{R}$-vector space.

\subsection{Exceptional Jordan algebra $\Vec{\mathfrak{J}}$}
We define $\Vec{\mathfrak{J}}$ \,as the exceptional Jordan algebra consisting of all $3 \times 3$ hermitian matrices $A$ \,with entries in the Graves-Cayley algebra $\Vec{\mathfrak{C}}$\,.
\begin{eqnarray}
\Vec{\mathfrak{J}} = \{ A \in M(3,\Vec{\mathfrak{C}}) \  | \  A^\ddagger = A \} \quad\quad \bigl( A^\ddagger \equiv (\bar{A})^T \bigr)
\end{eqnarray}
The specific components of $A$ \,can be written as follows.
\begin{eqnarray}
A
&=&
\left(
      \begin{array}{ccc}
      Q_1 & \phi_3 & \bar{\phi}_2 \\
      \bar{\phi}_3 & Q_2 & \phi_1 \\
      \phi_2 & \bar{\phi}_1 & Q_3
      \end{array}
\right) \\
& &
\quad\quad\quad Q_I \in \mathbf{R} \ \quad \phi_I \in \Vec{\mathfrak{C}} \quad\quad (I=1,2,3) \nonumber
\end{eqnarray}
Therefore, \,$\Vec{\mathfrak{J}}$ \,is a 27-dimensional $\mathbf{R}$-vector space.

\subsection{Complex exceptional Jordan algebra $\Vec{\mathfrak{J}^c}$}
Let $\Vec{\mathfrak{J}^c}$, called the complex exceptional Jordan algebra, be the complexification of $\Vec{\mathfrak{J}}$.
\begin{eqnarray}
\Vec{\mathfrak{J}^c} = \{ A + iB \  | \  A,B \in \Vec{\mathfrak{J}} \ , \  i^2=-1 \}
\end{eqnarray}
Therefore, the specific components of $X \in \Vec{\mathfrak{J}^c}$ \,can be written as follows.
\begin{eqnarray}
X
&=&
\left(
      \begin{array}{ccc}
      Q_1 & \phi_3 & \bar{\phi}_2 \\
      \bar{\phi}_3 & Q_2 & \phi_1 \\
      \phi_2 & \bar{\phi}_1 & Q_3
      \end{array}
\right)
+
i
\left(
      \begin{array}{ccc}
      P_1 & \pi_3 & \bar{\pi}_2 \\
      \bar{\pi}_3 & P_2 & \pi_1 \\
      \pi_2 & \bar{\pi}_1 & P_3
      \end{array}
\right) \\
& &
\quad\quad\quad Q_I, P_I \in \mathbf{R} \ \quad \phi_I, \pi_I \in \Vec{\mathfrak{C}} \quad\quad (I=1,2,3) \nonumber \\
&=&
\left(
      \begin{array}{ccc}
      x_1 & \xi_3 & \bar{\xi}_2 \\
      \bar{\xi}_3 & x_2 & \xi_1 \\
      \xi_2 & \bar{\xi}_1 & x_3
      \end{array}
\right) \\
& &
\quad\quad\quad x_I \in \mathbf{C} \ \quad \xi_I \in \Vec{\mathfrak{C}^c} \quad\quad (I=1,2,3) \nonumber \\
&\equiv& X(x,\xi) \\
& &
\qquad\qquad\qquad
\left\{
    \begin{array}{l}
    x_I = Q_I + i P_I \\
    \xi_I = \phi_I + i \pi_I \\
    \bar{\xi}_I = \bar{\phi}_I + i \bar{\pi}_I
    \end{array}
\right. 
\nonumber
\end{eqnarray}
Accordingly, we can also define this $\Vec{\mathfrak{J}^c}$ as 
\begin{eqnarray}
\Vec{\mathfrak{J}^c} = \{ X \in M(3,\Vec{\mathfrak{C}^c}) \  | \  X^\ddagger = X \} \quad\quad \bigl( X^\ddagger \equiv (\bar{X})^T \bigr) \  .
\end{eqnarray}

\subsection{Two kinds of hermitian adjoints}
In $\Vec{\mathfrak{C}^c}$, \,there exist two conjugations: \ One is the complex conjugation $(\cdots)^\ast$, \,another is the octonionic conjugation $\overline{(\cdots)}$\,. \,As a result, in $\Vec{\mathfrak{J}^c}$, \,there exist two kinds of hermitian adjoints.
\begin{eqnarray}
X^\dagger &\equiv& (X^\ast)^T \\
X^\ddagger &\equiv& (\bar{X})^T
\end{eqnarray}

\subsection{Operations}
Now, for any $X,Y,Z \in \Vec{\mathfrak{J}^c}$, \,each of the operations is defined as follows.
\begin{eqnarray}
X
=
\left(
      \begin{array}{ccc}
      x_1 & \xi_3 & \bar{\xi}_2 \\
      \bar{\xi}_3 & x_2 & \xi_1 \\
      \xi_2 & \bar{\xi}_1 & x_3
      \end{array}
\right) \  , \quad
Y
=
\left(
      \begin{array}{ccc}
      y_1 & \eta_3 & \bar{\eta}_2 \\
      \bar{\eta}_3 & y_2 & \eta_1 \\
      \eta_2 & \bar{\eta}_1 & y_3
      \end{array}
\right) \  , \quad
Z
=
\left(
      \begin{array}{ccc}
      z_1 & \zeta_3 & \bar{\zeta}_2 \\
      \bar{\zeta}_3 & z_2 & \zeta_1 \\
      \zeta_2 & \bar{\zeta}_1 & z_3
      \end{array}
\right) \nonumber \\
\end{eqnarray}

\subsubsection{trace \  $tr ( X )$}
\begin{eqnarray}
tr ( X ) &\equiv& x_1 + x_2 + x_3
\end{eqnarray}

\subsubsection{Jordan multiplication \  $X \circ Y$}
\begin{eqnarray}
X \circ Y &\equiv& \frac{1}{2} ( XY + YX ) \\
             &=& \frac{1}{2} \{ X , Y \} \\
             &=& Y \circ X
\end{eqnarray}

\subsubsection{inner product \  $( X , Y ) \  \in \mathbf{C}$}
\begin{eqnarray}
( X , Y ) &\equiv& tr ( X \circ Y ) \\
          &=& \frac{1}{2} tr ( XY ) + \frac{1}{2} tr ( YX ) \\
          &=& ( Y , X )
\end{eqnarray}

\subsubsection{hermitian product \  $\langle X , Y \rangle \  \in \mathbf{C}$}
\begin{eqnarray}
\langle X , Y \rangle &\equiv& ( X^\ast , Y ) \\
\vspace{1cm} \nonumber \\
\Bigl( \quad 0 &\le& \langle X , X \rangle \quad \in \quad \mathbf{R} \quad \Bigr)
\end{eqnarray}

\subsubsection{Freudenthal multiplication \  $X \times Y$}
\begin{eqnarray}
X \times Y &\equiv& X \circ Y - \frac{1}{2} tr ( X ) Y - \frac{1}{2} tr ( Y ) X + \frac{1}{2} tr ( X ) tr ( Y ) I - \frac{1}{2} ( X , Y ) I \\
&=& Y \times X \\
& &
\qquad\qquad\qquad\qquad\qquad\qquad\qquad I: \ \mbox{unit matrix} \nonumber
\end{eqnarray}

\subsubsection{trilinear form \  $tr ( X , Y , Z ) \  \in \mathbf{C}$}
\begin{eqnarray}
tr ( X , Y , Z ) &\equiv& ( X , Y \circ Z ) \\
                            &=& tr ( X \circ ( Y \circ Z ) ) \\
&=& \frac{1}{4} tr ( X(YZ) ) + \frac{1}{4} tr ( X(ZY) ) \nonumber \\
& & {} + \frac{1}{4} tr ( (YZ)X ) + \frac{1}{4} tr ( (ZY)X ) \\
&=& \frac{1}{2} ( X,YZ ) + \frac{1}{2} ( X,ZY ) \\
&=& tr ( Y , Z , X ) \  = \  tr ( Z , X , Y ) \\
&=& tr ( Z , Y , X ) \  = \  tr ( Y , X , Z ) \  = \  tr ( X , Z , Y ) \\
&=& ( X \circ Y , Z )
\end{eqnarray}

\subsubsection{cubic form \  $( X , Y , Z ) \  \in \mathbf{C}$}
\begin{eqnarray}
( X , Y , Z ) &\equiv& ( X , Y \times Z ) \\
                            &=& tr ( X \circ ( Y \times Z ) ) \\
&=& tr ( X , Y , Z ) \nonumber \\
& & {} - \frac{1}{2} tr ( X ) \  ( Y , Z ) - \frac{1}{2} tr ( Y ) \  ( Z , X ) - \frac{1}{2} tr ( Z ) \  ( X , Y ) \nonumber \\
& & {} + \frac{1}{2} tr ( X ) \  tr ( Y ) \  tr ( Z ) \\
&=& ( Y , Z , X ) \  = \  ( Z , X , Y ) \\
&=& ( Z , Y , X ) \  = \  ( Y , X , Z ) \  = \  ( X , Z , Y ) \\
&=& ( X \times Y , Z )
\end{eqnarray}

\subsubsection{determinant \  $det ( X ) \  \in \mathbf{C}$}
\begin{eqnarray}
det ( X ) &\equiv& \frac{1}{3} ( X , X , X ) \\
&=& \frac{1}{6} tr ( X(XX) ) + \frac{1}{6} tr ( (XX)X ) - \frac{1}{2} tr ( X^2 ) tr ( X ) + \frac{1}{6} {tr ( X )}^3
\end{eqnarray}

\subsubsection{cycle mapping \  $\mathcal{P} ( X )$}

\begin{eqnarray}
X &=&
\left(
      \begin{array}{ccc}
      x_1 & \xi_3 & \bar{\xi}_2 \\
      \bar{\xi}_3 & x_2 & \xi_1 \\
      \xi_2 & \bar{\xi}_1 & x_3
      \end{array}
\right)
\end{eqnarray}

For any $X \in \Vec{\mathfrak{J}^c}$, \,the cycle mapping $\mathcal{P} ( X )$ \,is defined by 
\begin{eqnarray}
\mathcal{P} ( X )
&\equiv&
\left(
      \begin{array}{ccc}
      x_2 & \xi_1 & \bar{\xi}_3 \\
      \bar{\xi}_1 & x_3 & \xi_2 \\
      \xi_3 & \bar{\xi}_2 & x_1
      \end{array}
\right) \  .
\end{eqnarray}
Namely, the cycle mapping is the cyclic permutation with respect to the indices $I = 1, 2, 3$\,. \,Therefore, we have 
\begin{eqnarray}
\mathcal{P}^2 ( X )
&=&
\left(
      \begin{array}{ccc}
      x_3 & \xi_2 & \bar{\xi}_1 \\
      \bar{\xi}_2 & x_1 & \xi_3 \\
      \xi_1 & \bar{\xi}_3 & x_2
      \end{array}
\right) \  , \\
& & {} \nonumber \\
\mathcal{P}^3 ( X ) &=& 1 \cdot X \  = \  X \  .
\end{eqnarray}

\subsection{Some helpful formulas on elements of $\Vec{\mathfrak{J}^c}$}
We can use the following formulas for any $X,Y,Z \in \Vec{\mathfrak{J}^c}$.
\begin{eqnarray}
I \circ I &=& I \\
I \circ X &=& X \\
I \times I &=& I \\
I \times X &=& \frac{1}{2} ( tr(X) I - X ) \\
(I,I) &=& 3 \\
(X,I) &=& tr(X,I,I) \  = \  (X,I,I) \  = \  tr(X) \\
(X,Y) &=& tr(X,Y,I) \\
(X,YZ) &=& (Y,ZX) \  = \  (Z,XY) \\
tr(X \times Y) &=& \frac{1}{2} tr(X) tr(Y) - \frac{1}{2} (X,Y) \\
X \circ (X \times X) &=& det(X) \  I \\
& &
\qquad\qquad\qquad\qquad I: \ \mbox{unit matrix} \nonumber
\end{eqnarray}

\subsection{Indication by components}
\begin{eqnarray}
X \circ X &=& XX \\
&=&
\left(
      \begin{array}{ccc}
      (x_1)^2+\xi_2\bar{\xi}_2+\xi_3\bar{\xi}_3 & \overline{\xi_1\xi_2}+(x_1+x_2)\xi_3 & \xi_3\xi_1+(x_3+x_1)\bar{\xi}_2 \\
      \xi_1\xi_2+(x_1+x_2)\bar{\xi}_3 & (x_2)^2+\xi_3\bar{\xi}_3+\xi_1\bar{\xi}_1 & \overline{\xi_2\xi_3}+(x_2+x_3)\xi_1 \\
      \overline{\xi_3\xi_1}+(x_3+x_1)\xi_2 & \xi_2\xi_3+(x_2+x_3)\bar{\xi}_1 & (x_3)^2+\xi_1\bar{\xi}_1+\xi_2\bar{\xi}_2
      \end{array}
\right) \nonumber \\
\\
X \times X &=& 
\left(
      \begin{array}{ccc}
      x_2x_3-\xi_1\bar{\xi}_1 & \overline{\xi_1\xi_2}-x_3\xi_3 & \xi_3\xi_1-x_2\bar{\xi}_2 \\
      \xi_1\xi_2-x_3\bar{\xi}_3 & x_3x_1-\xi_2\bar{\xi}_2 & \overline{\xi_2\xi_3}-x_1\xi_1 \\
      \overline{\xi_3\xi_1}-x_2\xi_2 & \xi_2\xi_3-x_1\bar{\xi}_1 & x_1x_2-\xi_3\bar{\xi}_3
      \end{array}
\right) \\
\vspace{1cm} \nonumber \\
tr ( X Y ) &=& \sum_{I=1}^{3} \Bigl( x_I y_I + ( \bar{\xi}_I \eta_I + \xi_I \bar{\eta}_I ) \Bigr) \\
\vspace{1cm} \nonumber \\
tr ( X ( Y Z ) ) &=& \sum_{I=1}^{3} \Bigl( \  x_I y_I z_I + x_I ( (\bar{\eta}_{I+1} \zeta_{I+1}) + (\eta_{I+2} \bar{\zeta}_{I+2}) ) \nonumber \\
& & {} + y_I ( (\xi_{I+1} \bar{\zeta}_{I+1}) + (\bar{\xi}_{I+2} \zeta_{I+2}) ) + z_I ( (\bar{\xi}_{I+1} \eta_{I+1}) + (\xi_{I+2} \bar{\eta}_{I+2}) ) \nonumber \\
& & {} + ( \xi_I (\eta_{I+1} \zeta_{I+2}) + \overline{(\zeta_{I+1} \eta_{I+2}) \xi_I} ) \  \Bigr) \\
\vspace{1cm} \nonumber \\
tr ( ( X Y ) Z ) &=& \sum_{I=1}^{3} \Bigl( \  x_I y_I z_I + x_I ( (\bar{\eta}_{I+1} \zeta_{I+1}) + (\eta_{I+2} \bar{\zeta}_{I+2}) ) \nonumber \\
& & {} + y_I ( (\xi_{I+1} \bar{\zeta}_{I+1}) + (\bar{\xi}_{I+2} \zeta_{I+2}) ) + z_I ( (\bar{\xi}_{I+1} \eta_{I+1}) + (\xi_{I+2} \bar{\eta}_{I+2}) ) \nonumber \\
& & {} + ( (\xi_I \eta_{I+1}) \zeta_{I+2} + \overline{\zeta_{I+1} (\eta_{I+2} \xi_I)} ) \  \Bigr) \\
\vspace{1cm} \nonumber \\
( X , Y ) &=& \sum_{I=1}^{3} \Bigl( x_I y_I + 2( \xi_I , \eta_I ) \Bigr) \\
\vspace{1cm} \nonumber \\
\langle X , Y \rangle &=& \sum_{I=1}^{3} \Bigl( x_I^\ast y_I + 2\langle \xi_I , \eta_I \rangle \Bigr) \\
\vspace{1cm} \nonumber \\
tr ( X , Y , Z ) &=& \sum_{I=1}^{3} \Bigl( \  x_I y_I z_I + x_I ( (\eta_{I+1},\zeta_{I+1}) + (\eta_{I+2},\zeta_{I+2}) ) \nonumber \\
& & {} + y_I ( (\zeta_{I+1},\xi_{I+1}) + (\zeta_{I+2},\xi_{I+2}) ) + z_I ( (\xi_{I+1},\eta_{I+1}) + (\xi_{I+2},\eta_{I+2}) ) \nonumber \\
& & {} + \Vec{Re^c} ( \xi_I \eta_{I+1} \zeta_{I+2} + \xi_I \zeta_{I+1} \eta_{I+2} ) \  \Bigr) \\
\vspace{1cm} \nonumber \\
( X , Y , Z ) &=& \sum_{I=1}^{3} \Bigl( \  \frac{1}{2} ( x_I y_{I+1} z_{I+2} + x_I y_{I+2} z_{I+1} ) - ( x_I (\eta_I,\zeta_I) + y_I (\zeta_I,\xi_I) + z_I (\xi_I,\eta_I) ) \nonumber \\
& & {} + \Vec{Re^c} ( \xi_I \eta_{I+1} \zeta_{I+2} + \xi_I \zeta_{I+1} \eta_{I+2} ) \  \Bigr)
\end{eqnarray}

Here, \,the index $I$ \,is $\Vec{mod \  3}$.

Therefore, we have 
\begin{eqnarray}
det ( X ) &=& x_1 x_2 x_3 - x_1 \xi_1 \bar{\xi}_1 - x_2 \xi_2 \bar{\xi}_2 - x_3 \xi_3 \bar{\xi}_3 + 2 \Vec{Re^c} ( \xi_1 \xi_2 \xi_3 ) \  .
\end{eqnarray}
%
%
%
%
%
%
                                                     %
%
%
%
%
%
%
%
%

\begin{thebibliography}{99}                                %
\bibitem{deWit:1988ig}
B.~de Wit, J.~Hoppe and H.~Nicolai,
{\itshape On the quantum mechanics of supermembranes},
Nucl.\ Phys.\ B {\bf 305} (1988), 545.
%
%
%
%
\bibitem{Banks:1997vh}
T.~Banks, W.~Fischler, S.~H.~Shenker and L.~Susskind,
{\itshape M theory as a matrix model: A conjecture},
Phys.\ Rev.\ D {\bf 55} (1997), 5112
[hep-th/9610043].
%
%
%
%
\bibitem{Ishibashi:1997xs}
N.~Ishibashi, H.~Kawai, Y.~Kitazawa and A.~Tsuchiya,
{\itshape A large-N reduced model as superstring},
Nucl.\ Phys.\ B {\bf 498} (1997), 467
[hep-th/9612115].
%
%
%
%
\bibitem{Dijkgraaf:1997vv}
R.~Dijkgraaf, E.~Verlinde and H.~Verlinde,
{\itshape Matrix string theory},
Nucl.\ Phys.\ B {\bf 500} (1997), 43
[hep-th/9703030].
%
%
%
%
\bibitem{Smolin:2000kc}
L.~Smolin,
{\itshape M theory as a matrix extension of Chern-Simons theory},
Nucl.\ Phys.\ B {\bf 591} (2000), 227
[hep-th/0002009].
%
%
%
%
\bibitem{Smolin:2000fr}
L.~Smolin,
{\itshape The cubic matrix model and a duality between strings and loops},
[hep-th/0006137].
%
%
%
%
\bibitem{Azuma:2001re}
T.~Azuma, S.~Iso, H.~Kawai and Y.~Ohwashi,
{\itshape Supermatrix models},
Nucl.\ Phys.\ B {\bf 610} (2001), 251
[hep-th/0102168].
%
%
%
%
\bibitem{Smolin:2001wc}
L.~Smolin,
{\itshape The exceptional Jordan algebra and the matrix string},
[hep-th/0104050].
%
%
%
%
\bibitem{Kim:2001up}
Bumsig Kim and Albert Schwarz,
{\itshape Formulation of M(atrix) model in terms of octonions},
(unpublished).
%
%
%
%
\bibitem{Chamseddine:1997zu}
A.~H.~Chamseddine and A.~Connes,
{\itshape The spectral action principle},
Commun.\ Math.\ Phys.\  {\bf 186} (1997), 731
[hep-th/9606001].
%
%
%
%
\bibitem{Klimcik:1998mg}
C.~Klim\v{c}\'{\i}k,
{\itshape Gauge theories on the noncommutative sphere},
Commun.\ Math.\ Phys.\  {\bf 199} (1998), 257
[hep-th/9710153].
%
%
%
%
\bibitem{Kawamoto:1999gi}
N.~Kawamoto,
{\itshape Non-string pursuit towards unified model on the lattice},
Prog.\ Theor.\ Phys.\ Suppl.\  {\bf 134} (1999), 84
[hep-th/9904089].
%
%
%
%
\bibitem{Ashtekar:1986yd}
A.~Ashtekar,
{\itshape New variables for classical and quantum gravity},
Phys.\ Rev.\ Lett.\  {\bf 57} (1986), 2244.
%
%
%
%
\bibitem{Ashtekar:1987gu}
A.~Ashtekar,
{\itshape New hamiltonian formulation of general relativity},
Phys.\ Rev.\ D {\bf 36} (1987), 1587.
%
%
%
%
\bibitem{Yokota:1990e6}
I.~Yokota,
{\itshape Realizations of involutive automorphisms $\sigma$ and $G^\sigma$ of exceptional linear Lie groups $G$, part $I$, $G = G_2$, $F_4$ and $E_6$},
Tsukuba J. Math. Vol.14 No.1 (1990), 185.
%
%
%
%
\bibitem{Yokota:1990e7}
I.~Yokota,
{\itshape Realizations of involutive automorphisms $\sigma$ and $G^\sigma$ of exceptional linear Lie groups $G$, part $II$, $G = E_7$},
Tsukuba J. Math. Vol.14 No.2 (1990), 379.
%
%
%
%
\bibitem{Yokota:1991e8}
I.~Yokota,
{\itshape Realizations of involutive automorphisms $\sigma$ and $G^\sigma$ of exceptional linear Lie groups $G$, part $III$, $G = E_8$},
Tsukuba J. Math. Vol.15 No.2 (1991), 301.
%
%
%
%
\bibitem{Kugo:1983bn}
T.~Kugo and P.~Townsend,
{\itshape Supersymmetry and the division algebras},
Nucl.\ Phys.\ B {\bf 221} (1983), 357.
%
%
%
%
\bibitem{Horowitz:2001gn}
G.~T.~Horowitz and L.~Susskind,
{\itshape Bosonic M theory},
J.\ Math.\ Phys.\  {\bf 42} (2001), 3152
[hep-th/0012037].
%
%
%
%
\bibitem{Freudenthal:1951gf}
Hans Freudenthal,
{\itshape Oktaven, ausnahmegruppen und oktavengeometrie},
Mathematisch Instituut der Rijksuniversiteit te Utrecht 1951.
%
%
%
%
\bibitem{Kodama:1988yf}
H.~Kodama,
{\itshape Specialization of Ashtekar's formalism to Bianchi cosmology},
Prog.\ Theor.\ Phys.\  {\bf 80} (1988), 1024.
%
%
%
%
\bibitem{Kodama:1990sc}
H.~Kodama,
{\itshape Holomorphic wave function of the universe},
Phys.\ Rev.\ D {\bf 42} (1990), 2548.
%
%
%
%
\bibitem{Smolin:1995qb}
L.~Smolin and C.~Soo,
{\itshape The Chern-Simons invariant as the natural time variable for classical and quantum cosmology},
Nucl.\ Phys.\ B {\bf 449} (1995), 289
[gr-qc/9405015].
%
%
%
%
%
\end{thebibliography}
\end{document}